# Three-Dimensional Modelling of Ionized Gas

## I. Did very massive stars of different metallicities drive the second cosmic reionization?

J. A. Weber[1], A. W. A. Pauldrach[1], J. S. Knogl[1], and T. L. Hoffmann[1]

Institut für Astronomie und Astrophysik der Universität München, Scheinerstraße 1, 81679 München, Germany
e-mail: jweber@usm.lmu.de, uh10107@usm.lmu.de, sebastian.knogl@tum.de, hoffmann@usm.lmu.de

**ABSTRACT**

*Context.* The first generation of stars, which formed directly from the primordial gas, is believed to have played a crucial role in the early phase of the epoch of reionization of the universe. Theoretical studies indicate that the initial mass function (IMF) of this first stellar population differs significantly from the present IMF, being top-heavy and thus allowing for the presence of supermassive stars with masses up to several thousand solar masses. The first generation of population III stars was therefore not only very luminous, but due to its lack of metals its emission of UV radiation considerably exceeded that of present stars. Because of the short lifetimes of these stars the metals produced in their cores were quickly returned to the environment, from which early population II stars with a different initial mass function and different spectral energy distributions (SEDs) were formed, already much earlier than the time at which the universe became completely reionized (at a redshift of $z \gtrsim 6$).
*Aims.* Using a state-of-the-art model atmosphere code we calculate realistic SEDs of very massive stars (VMSs) of different metallicities to serve as input for the 3-dimensional radiative transfer code we have developed to simulate the temporal evolution of the ionization of the inhomogeneous interstellar and intergalactic medium, using multiple stellar clusters as sources of ionizing radiation. The ultimate objective of these simulations is not only to quantify the processes which are believed to have lead to the reionized state of the universe, but also to determine possible observational diagnostics to constrain the nature of the ionizing sources.
*Methods.* The multi-frequency treatment in our combination of 3d radiative transfer – based on ray-tracing – and time-dependent simulation of the ionization structure of hydrogen and helium allows, in principle, to deduce information about the spectral characteristics of the first generations of stars and their interaction with the surrounding gas on various scales.
*Results.* As our tool can handle distributions of numerous radiative sources characterized by *high resolution synthetic SEDs*, and also yields occupation numbers of the required energy levels of the most important elements which are treated in NLTE and are calculated consistently with the 3d radiative transfer, the ionization state of an inhomogeneous gaseous density structure can be calculated accurately. We further demonstrate that the increasing metallicity of the radiative sources in the transition from population III stars to population II stars has a strong impact on the hardness of the emitted spectrum, and hence on the reionization history of helium.
*Conclusions.* A top-heavy stellar mass distribution characterized by VMSs forming in chemically evolved clusters of high core mass density may not only provide the progenitors of intermediate-mass and supermassive black holes (SMBHs), but also play an important role for the reionization of He II. The number of VMSs required to reionize He II by a redshift of $z \sim 2.5$ is astonishingly close to the number of VMSs required to explain galactic SMBHs if one assumes that these have been formed by mergers of smaller black holes.

## 1. Introduction

Electrons and protons combined to form neutral hydrogen when the young universe cooled (at a redshift of $z \approx 1100$), but the continuous absorption troughs blueward of the (redshifted) hydrogen Ly$\alpha$ line, that one would expect to see if the intergalactic medium consisted of neutral (recombined) hydrogen gas, were not found in the spectra of distant objects around $z \approx 2.0$ (Gunn & Peterson 1965). It turns out that one has to look much deeper to find these troughs, and by means of quasar observations, White et al. (2003) and Kashikawa (2007) were able to show that the troughs do exist for objects at a redshift of $z \approx 6.5$, but not for objects at a redshift of $z \approx 5.7$. Obviously, a process has occurred that re-ionized the intergalactic gas in the universe, and this process must have been completed by a redshift of roughly $z \approx 6$.[1] As a result of this process the number density of neutral hydrogen is now very low in the intergalactic medium (IGM). At which redshifts it was still high is, due to the lack of corresponding observations, more difficult to determine, and one currently has to rely on numerical cosmological simulations to estimate when exactly reionization began. The simulations of Wise & Abel (2008) point for instance to a redshift of roughly $z \approx 30$ as starting point of the reionization process and marking the end of the "cosmic dark ages".[2]

---

[1] Fan et al. (2006b) have observed quasars in the redshift range from $5.7 \leq z \leq 6.13$ and have been able to detect deep Gunn-Peterson absorption gaps in Ly$\alpha$ for two quasars ($z = 6.13$ and $z = 6.01$).

[2] The investigation of the star formation history as a crucial ingredient in reionization modelling requires the consideration of both the large-scale structure formation and the smaller-scale aspects of star formation. At the present stage such considerations are treated with a number of approximations, for example a "threshold density" which describes a certain value of the gas density above which star formation is assumed to occur. We note that a discussion required about the relationship of this threshold value and the onset of star formation can be found in Maio et al. (2009), who concluded that low threshold densities lead to an early onset of star formation ($z \approx 25\ldots31$), while for larger values the onset of star formation is delayed ($z \approx 12\ldots16$).

Regarding the nature of the sources that have driven reionization we are still left somewhat in the dark. Although it was immediately recognized that quasars, as the brightest single objects in the UV, should have played an important role in this process, a quantitative investigation of the luminosity function of high-redshift quasars showed that their radiation output was not sufficient to reionize the universe (Fan et al. 2006a).[3] The other likely candidate objects to have significantly driven the reionization process are stars, the first generation of which (so-called population III stars) are assumed to have developed in the dark matter halos that had already formed at those redshifts.

There is reason to believe that due to their primordial composition and the corresponding inefficient cooling of the gas during the star-formation process these population III stars formed on a larger fragmentation scale ($10^2 \ldots 10^3 M_\odot$) without subfragmenting further (cf. Abel et al. 2000, Bromm et al. 2002), and therefore the IMF of the primordial stars differs significantly from the IMF at present days with a preference to massive stars (Bromm et al. 2002 give masses of $100 \leq M/M_\odot \leq 1000$, and temperatures of $T_{\text{eff}} \sim 10^5$ K, which to a large extent turn out to be independent of the stellar masses for these short living ($\tau_{\text{MS}} \sim 2 \cdot 10^6$ yr, cf. Schaerer 2002) objects). Although the exact form of this top-heavy IMF is still under debate[4], the preference for massive stars leads to huge UV luminosities, and this energy – released at high frequencies – may have been crucial for the evolution of the universe in its early stages.

Thus, the existence and the properties of the first stellar population is closely linked to the reionization history of the universe. But the time period for the appearance of these objects was much shorter than the time required to reionize the IGM, and different stellar populations will have contributed to the reionization at different times. In this context the history of reionization may be divided into three stages. The first one is the "pre-overlap" phase which was characterized by isolated H II regions located in the interstellar medium (ISM) which expanded steadily into the low density IGM. The second one is the "overlap" phase in which the separate H II regions started to merge; this process required a strong rise in the amount of ionizing radiation. The last stage is the "post-overlap" phase, characterized by ionizing the low density IGM on large scales (Gnedin 2000).

A more detailed description of this history has been presented by Cen (2003) whose analysis indicated that the universe was reionized in two steps. The first one occurred up to a redshift of $z \sim 15$ where the ionization was primarily driven by population III stars. As the population III stars returned metals via supernova explosions into their environment, the chemical composition and – as a result – the IMF and SEDs of later populations of stars changed, giving way to metal-enriched population II stars.[5] This transition from population III to population II stars lead to a decline in the emitted radiation power by a factor of ~10, which in turn lead to a phase, from about $z \approx 13$ to $z \approx 6$, during which hydrogen partly recombined ($n_{\text{H\,II}}/n_{\text{H}} \geq 0.6$) (Cen 2003). Thus the universe may have come close to a second cosmological recombination at $z \sim 13$. To overcome the threatening re-recombination and to restart the reionization process driven by powerful ionizing sources and maintained by the dilution of the intergalactic gas due to the expansion of the universe, a second step (finishing at $z \sim 6$) was necessary.[6] But how could the presumably much less massive population II and population I stars fully reionize the hydrogen and helium content of the IGM? Of course, there is strong evidence of an increasing star formation rate (SFR) in this epoch (Lineweaver 2001, Barger et al. 2000), but in particular regarding the reionization of helium not only the *number* of stars is important, but also the characteristics of the SEDs of the stars are much more significant (Wyithe & Loeb 2003). Very massive stars (VMSs) which clearly exceed the assumed upper limit of 100 to 150 $M_\odot$ of present-day massive stars have pronounced fluxes in the EUV *at all metallicities* (Pauldrach et al. 2012) and in conjunction with a top-heavy stellar mass distribution may provide such a required SED.

To motivate this idea we note that there are indications of a more top-heavy IMF than the standard one even in present-day galaxies. Harayama et al. (2008) for instance studied the IMF of one of the most massive Galactic star-forming regions – NGC 3603 – to verify whether the IMF really does have a universal shape. What they found does not support this assumption: with $\Gamma = -0.74$ they derived a power-law index for this massive starburst cluster which is considerably less steep than the Salpeter IMF ($\Gamma = -1.35$). Their result thus supports the hypothesis that a top-heavy IMF is not unusual for massive star-forming clusters and starburst galaxies even at solar metallicity. They further argued that a common property among starbursts showing such flat IMFs would be the high stellar density in the core of the cluster, and that the variations of the IMF could be linked to the spatial density of the stellar population (according to their results the core mass density of NGC 3603 is at least $6 \cdot 10^4 M_\odot$ pc$^{-3}$).

Moreover, very massive stars are efficient emitters of ionizing photons with approximately 10 times more hydrogen-ionizing photons and 10 000 times more He II-ionizing photons per unit stellar mass than a population with a Salpeter IMF. Pauldrach et al. (2012) discuss a mechanism to form such objects in present-day stellar clusters. They combined consistent models of expanding atmospheres with stellar evolutionary calculations of massive single stars with regard to the evolution of dense stellar clusters, and investigated the conditions necessary to initiate a runaway collision merger which may lead to the formation of a very massive object (up to several 1000 $M_\odot$) at the center of the cluster, and the possible formation of intermediate-mass black holes (IMBHs).[7] Interestingly these theoretical models have been observationally supported by two studies which demonstrate the existence of VMSs in the local Universe. The first one was based on HST and VLT spectroscopy from which Crowther et al. (2010) concluded that the dense cluster R136 located in the 30 Doradus region of the Large Magellanic Cloud hosts several stars whose initial masses were up to $\sim 300\,M_\odot$. The second study concerned the discovery of an optical transient which was classified as a Type Ic supernova (SNF20070406-008). As the observed light curve of this object fits that of a pair-instability supernova with a helium core mass of at least $100\,M_\odot$, the progenitor of this supernova must have been a VMS (Gal-Yam et al. 2009). Runaway collision mergers clearly exceed the assumed upper limit of $100\ldots150\,M_\odot$ for the direct formation of present-day massive stars, and since such mergers can also occur in chemically evolved clusters of high core mass density, they may have played an important role at least in the late stages of the reionization history.

---

[3] From observations it became evident that, compared to the UV background provided by the stellar content of normal star-forming galaxies, quasars supplied a fraction of less than 30% of the total ionizing radiation required to ionize the IGM (Willott et al. 2005, Mahabal et al. 2005). If the luminosity function of high-redshift quasars turns out to diverge considerably from the one at $z < 3$ (Fan et al. 2001) this estimate will have to be re-examined, but it is considered unlikely because theoretical models indicate that quasars have been a minor contributor to the reionization process even at a redshift of $z \approx 6$ (Volonteri & Gnedin 2009).

[4] In contrast to other studies Clark et al. (2011) found a stronger fragmentation cascade. However, the products at the lower end of the fragmentation cascade should be observed in the local universe, and up to now this has not been the case. We note, however, that Caffau et al. (2012) reported the finding of a 0.8 $M_\odot$ star with a metallicity of only $Z \approx 5 \cdot 10^{-5} Z_\odot$. This means that such a low metallicity may already be sufficient to produce enough fragmentation driven by dust cooling (cf. Klessen et al. 2012 and Schneider et al. 2012) to yield an IMF significantly different from that in the real primordial gas.

[5] Maio et al. (2010) found that the relative population III star formation rate dropped to $SFR_{\text{popIII}}/SFR_{\text{tot}} \approx 10^{-3}$ at $z \approx 13 \ldots 10$. Thus, the transition from population III to population II stars already occurred somewhat earlier than this. We further note that, due to the radiative feedback in the environment of the first stars, the formation of so-called "population III.2" stars occured even in primordial gas (cf. Norman 2010).

[6] In contrast to the study of Cen (2003), Furlanetto & Loeb (2005) assumed a more continuous transition from population III to population II stars and therefore a more monotonic reionization history. Observations of the 21 cm line of neutral hydrogen with the LOFAR radio interferometer and the Square Kilometer Array (SKA) will allow tracing the expansion of the ionized regions in the redshift range $z = 11.5 \ldots 6.5$ (Jelic 2010 and Santos et al. 2011) and help clarify the situation.

This conclusion is relevant in particular for the reionization of He II to He III, since this process was not yet completed at a redshift of $z \sim 6$. (Spectroscopic observations of He II Lyman $\alpha$ absorption troughs of quasars can be observed even at $z = 2.8$; Reimers et al. 1997, Kriss et al. 2001, and Syphers et al. 2011.) The reionization of He II was therefore considerably delayed compared to the reionization of H I and He I. Whether the appearance of population II and population I VMSs could have been responsible for the reionization of He II has not yet been investigated; it is presently an open point whether these very special stars, or, as assumed by Wyithe & Loeb (2003), Gleser et al. (2005), and McQuinn et al. (2009), quasars, or a mixture of these objects have been responsible for the reionization of He II.

In order to simulate and understand the reionization process different numerical approaches have been developed for a description of the radiative transfer and the evolution of the ionization fronts (e.g. Mellema et al. 2006, Gnedin & Abel 2001, Ciardi et al. 2001, Nakamoto et al. 2001, Razoumov & Cardall 2005, Ritzerveld et al. 2003, Alvarez et al. 2006, Reynolds et al. 2009, Iliev et al. 2009, Trac & Cen 2007, McQuinn et al. 2007, and Wise & Abel 2011). Many of these codes are specialized for a particular task and do not attempt to provide a comprehensive description of the time-dependent ionization structure, including the detailed statistical "equilibrium" of all relevant elements. E.g., for the purpose of simulating the radiative transfer in the early universe it is usually sufficient to just include tabulated ionization and recombination rates for hydrogen and helium, which are modelled as simple two- or three-level systems, to neglect metals, and to assume simplified models for the spectral energy distribution of the sources – e.g., black-body radiators or power-law spectra described by a few discrete wavelengths specifying the fluxes in the ranges important for the particular simulations.

Our focus in contrast lies in a sophisticated description of the 3d radiative transfer with respect to high spectral resolution, in order to quantify the evolution of all relevant ionization structures accurately, in particular also with the intent of determining possible observational features, such as line strength ratios from different ions. Metal lines, for instance, could be used to reconstruct the history of metal enrichment and the characteristics of the spectra of the ionizing sources (see, for instance, Graziani et al. 2012) that might be used to discriminate between different scenarios and/or particular aspects of these scenarios. Although only H and He are treated in the present paper, our algorithm in principle allows the implementation of multiple levels for each ionization stage of the metals, and we are currently working on the inclusion of metals as described by Hoffmann et al. (2012). We note in this regard that while metal ions have no direct significance as *absorbers* affecting the radiation field and thus the ionization of hydrogen and helium, metal *cooling* has a significant influence on the temperature structure of the gas and therefore on the temperature-dependent recombination rates of H and He. Thus, the metallicity dependent temperature has an impact on the ionization structure of the most abundant elements (e.g., Osterbrock & Ferland 2006). But this procedure only makes sense if realistic SEDs of the sources driving the reionization of the universe are supplied – from population III to population I stars at various metallicities, temperatures, and masses. To this aim we use a sophisticated model atmosphere code based on a consistent treatment of the expanding atmospheres of massive and very massive stars (cf. Pauldrach et al. 2001 and Pauldrach et al. 2011; note that the winds of hot stars modify the SEDs of the ionizing radiation of the stars dramatically, cf. Pauldrach 1987 and Pauldrach et al. 1994). We are therefore able to calculate the required SEDs to be used as input for the 3d radiative transfer simulating the ionization structure of the ISM and IGM.

In the following we will first introduce the theoretical basis of our simulations covering the concept of 3d radiative transfer and the physical mechanisms which influence the ionization structure of the gas surrounding the sources of ionization. We will further present a numerical method to describe the temporal expansion of the ionization fronts and which considers multiple ionization sources (Sect. 2). In Sect. 3 we discuss important details of our numerical approach and present tests comparing our 3d results with those of analytical solutions, a radially symmetric radiative transfer code, and results from a comparison project initiated by Iliev et al. (2006). In Sect. 4 we present first applications of our 3d radiative transfer on showcase simulations of the reionization scenario. Although using simplified initial conditions, we will demonstrate the influence of realistic stellar spectra of massive stars at different metallicities on the ionization structure of the surrounding gas under conditions which correspond to those of the early universe, and we will focus on the different behavior of the expansion of the ionization fronts with respect to a homogeneous gas density and an inhomogeneous cosmological density structure. With regard to the He II reionization problem we further investigate the influence of different input spectra on the ionization structure of He III by performing a series of multi-source simulations. We interpret and summarize our results along with an outlook in Sect. 5.

## 2. Three-dimensional radiative transfer as a tool for the description of the epochs of reionization

On larger scales radiation is emitted from point sources and this radiation starts to propagate homogeneously and isotropically in all directions up to a certain distance where material of the environment of the sources is encountered in the form of a density structure, which may also involve inhomogeneities or fluctuations, and radiative processes start to have a disturbing influence on this well-suited behavior. Moreover, the radiation propagating from the different sources will overlap at specific locations and this produces a time and frequency dependent 3-dimensional pattern of radiation. The evolution of such a radiation field and all its variables – intensity, radiative flux, photon flux – is however well defined and at every point in space characterized by the *equation of radiative transfer*:

$$\left[\frac{1}{c}\frac{\partial}{\partial t} - \frac{H\nu}{c}\frac{\partial}{\partial \nu} + \hat{\boldsymbol{n}} \cdot \frac{1}{a}\boldsymbol{\nabla}\right] I_\nu(\boldsymbol{r}, \hat{\boldsymbol{n}}, t) = \eta_\nu - \chi_\nu I_\nu(\boldsymbol{r}, \hat{\boldsymbol{n}}, t) \quad (1)$$

---

[7] Mass segregation as a result of stellar dynamics in a dense young cluster, associated with core collapse and the formation of a runaway stellar collision process, was promoted by Portegies Zwart et al. (2004) to explain ultra-luminous X-ray sources (ULX).

This Boltzmann equation (cf. Gnedin & Ostriker 1997) describes the transport of radiative energy via a change of the intensity $I_\nu$ caused by the absorption coefficient $\chi_\nu \equiv \chi_\nu(\boldsymbol{r}, \hat{\boldsymbol{n}}, t) = \kappa_\nu(\boldsymbol{r}, \hat{\boldsymbol{n}}, t) + \frac{3H}{c}$ and the emissivity $\eta_\nu \equiv \eta_\nu(\boldsymbol{r}, \hat{\boldsymbol{n}}, t)$ as a function of position $\boldsymbol{r}$ (in comoving coordinates), direction $\hat{\boldsymbol{n}}$, and time $t$ for every frequency $\nu$. In this equation, $c$ is the speed of light, $a \equiv a(t)$ is the cosmological scale factor, $H \equiv H(t) = \dot{a}(t)/a(t)$ is the Hubble expansion rate, and $\kappa_\nu$ is the opacity; as $\kappa_\nu/(3H/c)$ is on the order of $10^4 \ldots 10^6$ in the relevant frequency range of a neutral gas with a density equal to the mean density of the universe in the considered redshift range ($0 \le z \le 15$), we assume $\chi_\nu = \kappa_\nu$ in the remaining part of this paper.

Neglecting the explicit time derivative term (for a rationale see Sect. 3.2) and the frequency derivative term (justified if the universe does not expand significantly before the corresponding photons are absorbed – cf. Petkova & Springel 2009 and Wise & Abel 2011) in Eq. 1, and making use of the identity $\hat{\boldsymbol{n}} \cdot \frac{1}{a}\boldsymbol{\nabla} = \frac{d}{ds}$, which involves the interpretation of $\hat{\boldsymbol{n}} \cdot \frac{1}{a}\boldsymbol{\nabla}$ as a directional derivative along a path element $s$, the equation of radiative transfer becomes for each ray to be considered

$$\frac{dI_\nu(s)}{ds} = \chi_\nu(s)(S_\nu(s) - I_\nu(s)), \quad (2)$$

(where, as usual, $S_\nu$ represents the source function, defined as the ratio of emissivity to opacity, $S_\nu = \eta_\nu/\chi_\nu$). Equation 2 can now be solved formally in an analytical way

$$I_\nu(s) = I_\nu(s_0(n)) e^{\tau_\nu(s_0(n))-\tau_\nu(s)} + \int_{s_0(n)}^{s} \eta_\nu(s') e^{\tau_\nu(s')-\tau_\nu(s)} ds'. \quad (3)$$

$s_0(n)$ corresponds to the starting point of each ray $n$, and $\tau_\nu(s) = \int_{s_0(n)}^{s} \chi_\nu(s') ds'$ is the usual dimensionless optical depth variable quantifying the absorption characteristics of the medium.[8] Numerically the equation is solved using a suitable discretization scheme dividing the volume into small cells. Since by definition of this discretization scheme the opacity $\chi_\nu(s)$ stays constant along the cell-crossing distance $l_n(m)$ of each specific cell $m$ crossed by the considered ray $n$, the corresponding optical depth $\tau$ can be expressed as

$$\tau_\nu(s_n(m)) = \sum_{m'=m(s_0(n))}^{m(s)} \chi_\nu(m') l_n(m') \quad (4)$$

(where $m(s_0(n))$ denotes the cell where the corresponding source of the considered ray $n$ is placed and $m(s)$ denotes the cell which has just been crossed before $s$ is reached) and Eq. 3 thus simplifies to

$$I_\nu(s_n(m)) = I_\nu(s_0(n)) e^{-\tau_\nu(s_n(m))}. \quad (5)$$

It is now straightforward to calculate from Eq. 5 the energy

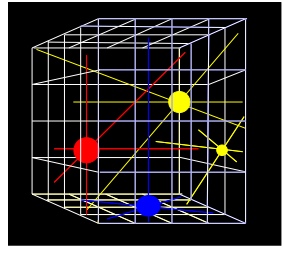

**Fig. 1.** Sketch showing a subset of the rays to be treated in the ray tracing method of radiative transfer. All rays start from the sources of radiation (filled dots), but the lifetime of the sources may be much shorter than the ionization time of the volume. Possible different spectral energy distributions of the sources are indicated by different colors.

deposited per time in the cells that are crossed by a ray $n$ over a distance $s_n(m)$

$$\Delta \dot{E}(s_n(m)) = \int_0^\infty \tilde{L}_\nu(s_0(n)) \left(1 - e^{-\tau_\nu(s_n(m))}\right) d\nu. \quad (6)$$

Here $\tilde{L}_\nu = L_\nu/N_{\text{rays}}$ is the fraction of the (spectral) luminosity emitted by the source into each of the $N_{\text{rays}}$ rays considered per source. (All rays start at sources (cf. Fig. 1), which represent single stars or star clusters, each with a specified spectral energy distribution $F_\nu(M_n, Z_n, T_{\text{eff}}(n), R_n)$ and spectral luminosity $L_\nu(s_0(n)) = 4\pi R_n^2 F_\nu(n)$, characterized by the source's mass $M_n$, metallicity $Z_n$, effective temperature $T_{\text{eff}}(n)$, and radius $R_n$ (for single stars, the stellar radius; for clusters, an equivalent radius describing the total radiating surface) – cf. Sect. 3.2).

### 2.1. 3-dimensional radiative transfer based on ray tracing

Because the equation of radiative transfer has to be solved explicitly for every time step at every point in space, a suitable discretization scheme is required that must be able to deal with possible inhomogeneities in the density distribution of the gas in the considered volume and/or the presence of a variable number of different radiative sources which in general will not possess any kind of symmetry (cf. Figs. 1 and 2). The natural property of light to propagate along straight lines straightforwardly leads to the concept of a *ray by ray solution* which involves distributing rays isotropically around each source and solving the transfer equation for each of these rays, taking into account the interaction of the gas and the photons along the way.[9]

The ray tracing concept along with a spatial discretization scheme described by a Cartesian coordinate system can naturally account for the following three important points:

---

[8] In applications of the IGM the diffuse radiation field (resulting from the emission term in Eq. 3) is usually not explicitly treated. We employ the commonly used "on-the-spot approximation" (Zanstra 1931, Baker & Menzel 1938, Spitzer 1998, cf. also Sect. 2.2.1) at the present stage, and quantify its influence on the photoionization state in Sect. 3.3.3.

[9] Although this ray tracing method appears natural for solving the 3-dimensional radiative transfer, a number of other authors have considered different approaches as well. Monte-Carlo techniques, where the paths of the rays are not computed in advance, but random walks of the photons are assumed instead, are often used (e.g., Ciardi et al. 2001, Ercolano et al. 2003). The primary advantage of this procedure is that it allows scattering and re-emission processes to be included straightforwardly, but its disadvantage is that an extremely large number of photon packets has to be considered to reduce the inherent random noise (cf. Fig. 8). A technique that does not involve individual rays is the "moment method" in which the radiative transfer is treated as a diffusion process (e.g., Nakamoto et al. 2001, Petkova & Springel 2009) and which allows the radiative transfer to be solved very quickly even for a large number of ionizing sources. The drawback of this technique, however, is a low geometric accuracy, in particular with regard to shadowing effects that naturally occur where inhomogeneous density distributions are involved (cf. Fig. 2). The basic requirements for an adequate solution of the 3-dimensional radiative transfer therefore lead us and others (Trac & Cen 2007, McQuinn et al. 2007, Wise & Abel 2011) to choose the ray tracing concept for their purposes.

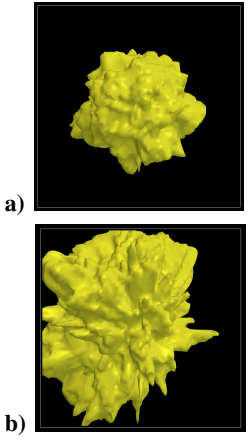

**Fig. 2.** Calculated ionization fronts for volumes of gas with inhomogeneous, fractal density structures. As the gas exposed to the ionizing radiation does not exhibit an intrinsic symmetry, a three-dimensional treatment of the radiative transfer is indispensable (for computational details see Sect. 3). In image a) the deviations from a homogeneous density distribution are much smaller than in image b), and consequently the shape of the ionized volume is less spherical in b) than in a). (The calculation of the fractal density distributions has been based on the algorithms presented by Elmegreen & Falgarone 1996 and Wood et al. 2005.)

- a radiation field that is generated by numerous different point sources arbitrarily distributed in space and characterized by individually different spectral energy distributions,
- an inhomogeneous density distribution of the medium,
- the temporal evolution of the radiation field and the propagation of the ionization fronts.

Although in our discretization scheme the physical conditions are treated as being spatially constant within each cell, and we consider (for reasons of efficiency) each of the emitting sources to be located in the center of its respective cell, we can account for arbitrary distributions of the gas and the sources with a resolution limited only by the computational resources. Time-dependencies can in principle be described accurately with a suitably fine resolution in space and time. With regard to the spatial resolution we note that it is also essential that the radiation field of each source is itself discretized with a matching angular resolution so that each cell is crossed by a sufficient number of rays in order to describe the effects of the radiation on the gas correctly (cf. Fig. 3 and Sect. 3.1).

### 2.2. Physical processes affecting the state of ionization

The basis of any approach in constructing detailed radiative models for the ionization structure of the gas surrounding clusters of sources of ionization is a concept that includes the time-dependent statistical equilibrium for all important ions with detailed atomic physics – described by rate equations –, the energy equation, and the radiative transfer equation at all transition frequencies required. As all of the involved equations have to be solved simultaneously, and as the replication of the required physical processes, which due to the energy input of the time varying sources continuously modify the physical properties of the surrounding gas, makes the solution of every realistic approach to a formidable problem, the method is not simple at all. We will in the following therefore give a description of the physics to be treated in some detail.

#### 2.2.1. The rate equations

As an essential step of the procedure the time-dependent occupation numbers $n \equiv n(r, t)$ of all ionization stages $i, j$ of the elements considered have to be calculated. The time-dependent statistical equilibrium

$$\frac{d}{dt}n_i = \sum_{i \neq j} \mathcal{P}_{j,i} n_j - \sum_{i \neq j} \mathcal{P}_{i,j} n_i, \qquad (7)$$

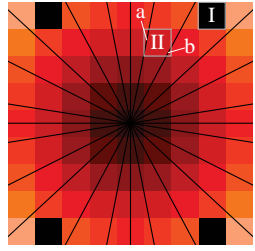

**Fig. 3.** Illustration of the spatial discretization scheme by example of a 2D layer containing $9 \times 9$ cells. The different physical conditions in each cell are represented by the different cell colors, and the black lines indicate the rays used in the radiative transfer. Insufficient ray densities where some cells are not crossed by any ray (e.g., cell I) must be avoided as this will lead to spurious results since the corresponding cells will not feel the effects of the radiation field.

which describes the temporal derivation of the number density of an ionization stage $i$, and which contains via the rate coefficient $\mathcal{P}_{i,j}$ all important radiative $\mathcal{R}_{i,j}$ and collisional $C_{i,j}$ transition rates, serves as a basis for this step[10]. In order to solve these systems of differential equations we define a vector $\mathbf{n}$, which contains the number densities of all ionization stages of the considered element, and a tridiagonal matrix $\mathbf{G}$, which contains in its components $g_{i,j}$ the rate coefficients[11]. With these definitions Eq. 7 is rewritten as

$$\mathbf{E} \cdot \frac{d}{dt}\mathbf{n} + \mathbf{G} \cdot \mathbf{n} = \mathbf{0}, \qquad (8)$$

where for reasons which will become obvious below the unity matrix $\mathbf{E}$ has been applied on the time derivative term on the left hand side of the system of equations. By scaling Eq. 8 with the total number density of the element considered, and by replacing the fraction $x_k$ of the ionization stage $k$ by the condition of particle conservation

$$x_k = 1 - \sum_{j, j \neq k}^{N} x_j, \qquad (9)$$

where $N$ is the number of ionization stages and $\mathbf{x}$ contains in its components the number densities of the ions relative to the total number density of the element, one gets the final system of inhomogeneous equations

$$\mathbf{E}' \cdot \frac{d}{dt}\mathbf{x} + \mathbf{G}' \cdot \mathbf{x} = \mathbf{b}, \qquad (10)$$

where the components of $\mathbf{G}'$ are given by $g'_{i,j} = g_{i,j} - g_{i,k}$, and those of $\mathbf{b}$ are $b_i = -g_{i,k}$, and where all coefficients of the redundant $k$-th row of $\mathbf{G}'$ and $\mathbf{b}$ have been replaced by 1, and those of the $k$-th column of $\mathbf{E}$ by 0 (with this replacement the unity matrix $\mathbf{E}$ becomes $\mathbf{E}'$) – with these numbers inserted the corresponding components represent in total the condition of particle conservation.

**Modelling the temporal evolution of the ionization structure.** The solution of the time-dependent rate equations (Eq. 10) required for modelling the temporal evolution of the ionization structures is not straightforward, because this system forms a set of stiff differential equations, and this means that the occupation numbers can change by several orders of magnitude for a gradual increase of the timescale[12]. On the other hand, for constant coefficients $\mathcal{P}_{i,j}$[13] the rate equations form a set of linear differential equations, and such a system can in principle be solved by an eigenvalue approach. With respect to this approach the general solution for the differential equations is based on the time independent "equilibrium" solution $x^\infty$ of the inhomogeneous system

$$\mathbf{G}' \cdot \mathbf{x}^\infty = \mathbf{b} , \quad (11)$$

and the solution of the homogeneous system

$$\mathbf{E}' \cdot \frac{\mathrm{d}}{\mathrm{d}t} \mathbf{x}_{hom}(t) + \mathbf{G}' \mathbf{x}_{hom}(t) = \mathbf{0} , \quad (12)$$

which result in the composed solution

$$\mathbf{x}(t) = \mathbf{x}_{hom}(t) + \mathbf{x}^\infty \quad (13)$$

of Eq. 10.

---

[10] As just processes which can alter the charge of an ion by ±1 are considered here, the net change $\frac{d}{dt}n_i$ of the occupation number of an ion $i$ simplifies to
$$\underbrace{n_{i+1}(\mathcal{R}_{i+1,i} + C_{i+1,i})}_{\mathcal{P}_{i+1,i} n_{i+1}} + \underbrace{n_{i-1}(\mathcal{R}_{i-1,i} + C_{i-1,i})}_{\mathcal{P}_{i-1,i} n_{i-1}} - \underbrace{n_i(\mathcal{R}_{i,i+1} + C_{i,i+1} + \mathcal{R}_{i,i-1} + C_{i,i-1})}_{\mathcal{P}_{i,i+1} n_i + \mathcal{P}_{i,i-1} n_i}$$
Note that the number densities $n_{l,u}$ of specific atomic levels $l, u$ of the ionization stages are not explicitly required at this stage – cf. Sect. 2.2.2 – and that the advection term has been neglected.

[11] $g_{i,i-1} = -\mathcal{P}_{i-1,i}$; $g_{i,i} = \mathcal{P}_{i,i-1} + \mathcal{P}_{i,i+1}$; $g_{i,i+1} = -\mathcal{P}_{i+1,i}$.

While the last component $\mathbf{x}^\infty$ of the composed vector is obtained by a direct solution of Eq. 11, the more interesting homogeneous component $\mathbf{x}_{hom}(t)$, which contains in its coefficients just difference values, can not always be determined straightly; its physical behavior however results directly from the structure of Eq. 12

$$\mathbf{x}_{hom}(t) = (\mathbf{x}(t_k) - \mathbf{x}^\infty) \cdot e^{-\delta(t-t_k)} , \quad (14)$$

where $t_k$ marks the begin of the considered timestep, and $\delta$ is the eigenvalue of the system[14].

The numerical solution of the time-dependent behavior of the ionization structures is on basis of the eigenvalue approach stable, even for large time steps (cf. Sect. 3.3). Nevertheless is an appropriate choice of the timestep sizes extremely important, since the evolution of the ionization structures has to be traced with a sufficient temporal accuracy. In the frame of our approach it turned out that the timestep-size can be controlled quite well by the following extrapolation method, which limits the maximal relative change of the ionization state within one timestep,

$$\Delta t_k = \Delta t_{k-1} \min_{\text{all cells } m} \left( f \cdot \left| \frac{n_{H_1}(m, t_{k-1})}{n_{H_1}(m, t_{k-1}) - n_{H_1}(m, t_{k-2})} \right| \right) , \quad (15)$$

where $\Delta t_{k-1}$ is the duration of the previous timestep, and $n_{H_1}(m, t_{k-2})$ and $n_{H_1}(m, t_{k-1})$ are the number densities of neutral hydrogen at the beginning of the previous and the current time steps, respectively[15].

An adequate description of the evolution of the ionization structures furthermore requires the consideration of the cosmological expansion for every timestep via a scaling of the path elements $\Delta s_n(m, t)$ described by $\hat{\mathbf{n}} \cdot \frac{1}{a(t)} \nabla = \frac{\mathrm{d}}{\mathrm{d}s(t)}$ (cf. Eq. 1), i.e., the correct cosmological scale factor $a(t)$ has to be known for every timestep of the simulation. For the redshift range of our simulations the scale factor can be approximated quite well by the formula

$$a(t) = \left( \frac{\Omega_{m,0}}{\Omega_{\Lambda,0}} \right)^{1/3} \sinh^{2/3} \left( \frac{3}{2} H_0 \, t \, \sqrt{\Omega_{\Lambda,0}} \right), \quad (16)$$

where $H_0$ is the Hubble constant at $z = 0$, and $\Omega_{m,0}$ and $\Omega_{\Lambda,0}$ are the usual relative contributions of matter and dark energy to the total energy of the present universe (cf. Carroll & Ostlie 2006). This is used to recompute all path elements $\Delta s_n(m, t)$ for each timestep $t_k$ according to

$$\Delta s_n(m, t_k) = \Delta s_n(m, t_{k-1}) \cdot \frac{a(t_k)}{a(t_{k-1})} \quad \forall n \forall m. \quad (17)$$

The particle densities $n(m, t_k)$ giving rise to the emission and absorption coefficients in the radiative transfer equation are expressed in units of the volumes $\Delta V(m, t_k)$ of the grid cells, which scale with the cube of the same factor $a(t_k)/a(t_{k-1})$, have to be rescaled analogously via

$$n(m, t_k) = n(m, t_{k-1}) \cdot \frac{\Delta V(m, t_{k-1})}{\Delta V(m, t_k)} = n(m, t_{k-1}) \cdot \left( \frac{a(t_{k-1})}{a(t_k)} \right)^3 \forall m. \quad (18)$$

### 2.2.2. The rate coefficients

In the following paragraphs we describe briefly the ionization and recombination rates we used as coefficients for the rate matrix $\mathbf{G}$. Under nebular conditions, only ionization from the ground state will be important, whereas recombination may occur also to excited levels.[16] Therefore the total rates between two ionization stages are given by the sums of the rate coefficients of the specific atomic energy levels $l$ – ground state or excited level of the lower ionization stage $i$ – and $u$ – ground state or excited level of the upper ionization stage $i + 1$

$$n_i \mathcal{P}_{i,j} = \sum_{\substack{l \in i \\ u \in i+1}} n_l P_{lu} , \quad n_j \mathcal{P}_{j,i} = \sum_{\substack{u \in i+1 \\ l \in i}} n_u P_{ul} , \quad (19)$$

where $P_{lu}$ and $P_{ul}$ are the rate coefficients connecting these levels, and which contain all important radiative ($R_{ul}$, $R_{lu}$) and collisional ($C_{ul}$, $C_{lu}$) transition rates.

---

[12] This is of course not the case for those parts of the medium, which are still recombined or are already completely ionized – these parts are close to their respective statistical equilibriums $\mathbf{x}^\infty$ –; it is however obvious that time dependent drastic changes of the ionization fractions occur, if the ionization fronts reach the material, and since the ionization fronts of different ions are furthermore located at different spatial positions for every timestep (cf. Sect. 2.1 and Sect. 3.1.3), the application of standard approaches – like the methods by Euler and Runge-Kutta which represent explicit methods, or the method by Rosenbrock (cf. Press et al. 1992) which represents an implicit method – is extremely problematic in order to find an accurate solution for the differential equations.

[13] We note that the rate coefficients actually depend on the radiation field and the electron density of each cell, which in turn are functions of the simulated ionization structure. To assume that the rate coefficients can be kept constant over a timestep is nevertheless a reasonable approximation, since this assumption is consistent with the description of the physical state of the cells, given that the same assumption has been applied for the spatial discretization scheme.

[14] For elements which have only two or three ionization stages an analytical solution for the eigenvalue $\delta$ has been found, and for the case of a two state system this eigenvalue is simply described by $\delta = \mathcal{P}_{1,2} + \mathcal{P}_{2,1}$ (cf. Schmidt-Voigt & Koeppen 1987).

[15] The interval of the new timestep is determined on basis of the biggest change of the fraction of neutral hydrogen during the previous timestep and the parameter $f$ is used to adjust this value. Small values of $f$ obviously lead to shorter timesteps, and these are required for a correct description of the temporal evolution of the ionization structure, whereas large values reduce the number of necessary iterations in cases where just a stationary solution is of interest (the values chosen for $f$ range from 0.1 to 1.0; in Sect. 3.3 it is shown how $f$ affects the accuracy of the simulations).

[16] Thus, transitions between the levels must also be considered, to allow cascading to the ground level. The rate coefficients for these line transitions from level $l$ to level $l'$ are given by $R_{ll'} = A_{ll'}$ where $A_{ll'}$ are the Einstein coefficients of the transitions.

Photoionization and computation of the mean intensity. The photoionization rates are obtained by performing the frequency integral over the mean intensity $J_\nu$ weighted by the frequency-dependent ionization cross-section $\alpha_{lu}(\nu)$

$$R_{lu} = \int_{\nu_{lu}}^{\infty} \frac{4\pi\alpha_{lu}(\nu)}{h\nu} J_\nu \, d\nu, \qquad (20)$$

where $\nu_{lu}$ denotes the threshold frequency for the corresponding ionization edge. The computation of the ionization structures thus requires the calculation of the mean intensities $J_\nu$ for every grid cell. As these physical quantities cannot simply be calculated as numerical integrals of the radiative intensities over the entire sphere – $4\pi J_\nu = \oint I_\nu \, d\omega$, here $d\omega$ represents the solid angle –, as is the case for radiative transfer models which are based on plane-parallel or spherically symmetric geometries, a recipe for the evaluation of the $J_\nu$ values, which takes into account the discrete nature of the rays, must be provided. This recipe requires on the one hand knowledge of the function value $\dot{N}_\nu$, which describes the number of photons transported per time and frequency by a ray $n$ to the edge of a considered cell $m$ positioned at a distance $s_n(m)$ from the starting point of the ray (cf. Sect. 2)

$$\dot{N}_\nu(s_n(m)) = \frac{\tilde{L}_\nu(s_0(n)) \, e^{-\tau_\nu(s_n(m))}}{h\nu} \qquad (21)$$

(note that this equation results directly from the considerations of Eq. 4 and Eq. 6), and the function value $\Delta\dot{N}_\nu(m,n)$ of the number of photons absorbed in the considered cell from ray $n$ via the ray segment $l_n(m)$ is analogously given by

$$\Delta\dot{N}_\nu(m,n) = \dot{N}_\nu(s_n(m)) \left(1 - e^{-\chi_\nu(m) l_n(m)}\right). \qquad (22)$$

Summing over all rays crossing the considered cell by ray segments finally gives the function value $\Delta\dot{N}_\nu(m)$ of the total number of photons absorbed in the cell

$$\Delta\dot{N}_\nu(m) = \sum_{\substack{\text{ray} \\ \text{segments } n(m)}} \dot{N}_\nu(s_n(m)) \left(1 - e^{-\chi_\nu(m) l_n(m)}\right). \qquad (23)$$

The recipe requires on the other hand also knowledge about the total number of radiative processes occurring per time and frequency unit in a cell. This number, the number of the sum of absorbing particles in a cell – given by the product of the number densities of potential absorbers and their cross-sections, which defines as a sum the opacity $\chi_\nu(m)$, times the volume $V$ of the cell times the number of available photons per time and frequency unit in the cell –, is represented by the function value[17]

$$\Delta\dot{R}_\nu(m) = \chi_\nu(m) \cdot V \cdot \frac{4\pi J_\nu(m)}{h\nu}. \qquad (24)$$

The recipe is thus based on the fact that the number of radiative processes consuming photons in a cell has to be equal to the number of photons absorbed along the ray segments in that cell. That is, $\Delta\dot{R}_\nu(m) = \Delta\dot{N}_\nu(m)$ and the consistent value of the mean intensity $J_\nu$ is obtained from Eq. 23 and Eq. 24

$$J_\nu(m) = \sum_{\substack{\text{ray} \\ \text{segments } n(m)}} \frac{\tilde{L}_\nu(s_0(n)) \, e^{-\tau_\nu(s_n(m))} \left[1 - e^{-\chi_\nu(m) l_n(m)}\right]}{4\pi \chi_\nu(m) \, V}. \qquad (25)$$

---

[17] For example, if only photoionization by one type of particle is considered as opacity source, then $\chi_\nu(m) = n_l(m)\alpha_{lu}(\nu)$, and $\Delta\dot{R}_\nu(m)$ is given by $\Delta\dot{R}_\nu(m) = V \cdot n_l(m) \cdot \frac{4\pi\alpha_{lu}(\nu)}{h\nu} J_\nu(m)$, as it must be to be consistent with Eqs. 19, 20, and 24.

Radiative recombination. The radiative recombination rate coefficients are computed analogously to the photoionization rates

$$R_{ul} = \left(\frac{n_l}{n_u}\right)^* \int_{\nu_{lu}}^{\infty} \frac{4\pi\alpha_{lu}(\nu)}{h\nu} \left(\frac{2h\nu^3}{c^2} + J_\nu\right) e^{-h\nu/kT(m)} d\nu, \qquad (26)$$

where $k$ is the Boltzmann constant, $T(m)$ is the local value of the temperature stratification of the gas, and $\left(\frac{n_l}{n_u}\right)^*$ is the Saha-Boltzmann factor – the ratio of the occupation numbers that would be reached in the case of thermodynamic equilibrium.[18]

### 2.2.3. The temperature structure

As the recombination rates (but also emission line intensities and other diagnostic observables) depend directly on the temperature (cf. Eq. 26), accurate ionization structure modelling requires simultaneously determining the local temperature at every point in the gas. The physical basis for this is the microscopic energy equation which, in principle, states that the energy going into and coming out of the material must be conserved. The standard method for determining the temperature is balancing the energy gains and losses to the electron gas, including all processes that affect the electron temperature: bound-free transitions (ionization and recombination), free-free transitions, and inelastic collisions with ions. Pauldrach et al. (2001) and Hoffmann et al. (2012) give an overview of how the considered heating and cooling rates are computed in our simulations. In contrast to the method described there, however, here we do not restrict our computations to the stationary case, but instead account for temporal changes of the energy content in a grid cell with a time-dependent approach[19] in parallel with the time-dependent computation of the ionization structure (cf. Sect. 2.2.1). At the present stage we consider the ionization of only hydrogen and helium and therefore neglect cooling processes by metals (e.g., collisional cooling by the O III ion) which are dominant for metallicities characteristic of population II and population I stars. In models which simulate an already partly metal-enriched universe we therefore approximate the temperature from case to case with plausible values.

### 2.2.4. The spectral energy distribution of the radiation field

The simulation of the evolution of the ionization structures – obtained via an iteration of the radiative transfer and the rate equations – requires a representative set of frequency points at which the radiative transfer equation has to be solved. The total number of frequency points which has to be considered in this regard depends on the atomic physics which determines the distribu-

---

[18] As we apply the so called "on-the-spot approximation" in the present stage, the total radiative recombination rate $\mathcal{R}_{j,i}$ has to be reduced by the process of direct recombination to the ground state of the lower ionization stage (in the frame of this approximation it is assumed that photons which are emitted by recombinations to the ground state are locally re-absorbed – case-B approximation). For large gas temperatures ($T > 10\,000$ K) collisions between atoms/ions and electrons are also important processes, we thus use for the collisional ionization rates an approximative formula – cf. (Mihalas 1978, p. 134) – and the collisional recombination rates are computed by multiplying these ionization rates with the Saha-Boltzmann factor.

[19] For the required iterative procedure we use a linearized "Euler-method" to extrapolate a temperature that balances the heating and cooling rates along with the temporal changes.

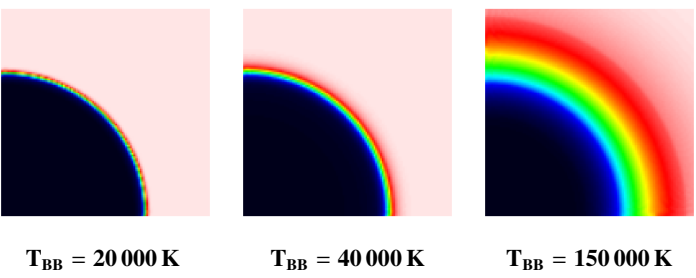

$T_{BB} = 20\,000$ K    $T_{BB} = 40\,000$ K    $T_{BB} = 150\,000$ K

**Fig. 4.** Calculated hydrogen equilibrium ionization structure as a function of the spectral energy distribution of the ionizing source, exemplified by black bodies with different temperatures but the same number of ionizing photons. Due to smaller values of the absorption cross section for photons with higher energies the shell of the "ionization front" becomes thicker for harder ionization spectra. This well-known effect can be reproduced only by models with a sufficient number of frequency points in the ionization continuum. The first panel corresponds to the result one usually obtains if the gas is illuminated by a monochromatic source.

**Table 1.** The 100 frequency points presently treated in our 3d radiative transfer calculations have been distributed over the important energy intervals as follows:

| Energy Interval | No. of frequency points |
|---|---|
| $h\nu < E_{\mathrm{ion,H\,I}}$ | 32 |
| $E_{\mathrm{ion,H\,I}} \leq h\nu < E_{\mathrm{ion,He\,I}}$ | 20 |
| $E_{\mathrm{ion,He\,I}} \leq h\nu < E_{\mathrm{ion,He\,II}}$ | 26 |
| $E_{\mathrm{ion,He\,II}} \leq h\nu$ | 22 |

tion of the frequency points over the relevant spectral range[20]. As only hydrogen and helium have been treated as elements for the simulations presented in this paper, just hundred frequency points had yet to be considered. Table 1 shows the distribution of the required points over the relevant spectral range, and Fig. 4 demonstrates the effect of the wavelength-dependent absorption cross-section on the transition of neutral to ionized hydrogen. The images in the figure show the ionization structure obtained for blackbody sources of different temperatures, but the same number of hydrogen-ionizing photons per second ($10^{49}$).[21] For a soft ionizing spectrum (e.g. $T_{\mathrm{eff}} = 20\,000$ K) most of the ionizing photons have energies slightly above the Lyman-edge, and because of the large value of the ionization cross section at this frequency, the ionization front is displayed by a thin shell (this is the result one usually obtains if the gas is illuminated by a monochromatic source). In contrast, a hard ionizing spectrum (e.g. $T_{\mathrm{eff}} = 150\,000$ K) produces a mean energy of the photons which is considerably above the Lyman-edge, and since the ionization cross section has smaller values at such frequencies, the photons penetrate deeper into the gas – because of their larger mean free paths

$$\lambda_\nu = \frac{1}{\sum_i \chi_i(\nu)} \, , \qquad (27)$$

(where the total opacity $\chi_\nu$ has contributions $\chi_i(\nu)$ from each type of absorbing atom/ion $i$) – and the "ionization front" is therefore displayed by a thick shell. *This well-known effect can be reproduced only by models with a sufficient number of frequency points in the ionization continuum.*

## 3. Implementation and optimization details of the algorithm

The physical equations described in Sect. 2 are the guiding principles for the algorithm used to calculate the radiative transfer in a homogeneous or inhomogeneous gaseous medium. The numerical procedures which realize this algorithm are based on an iterative solution of the time-dependent radiative transfer (cf. Sect. 3.2) and the time dependent statistical description of the microphysical state of the gas, which together characterize the ionization structures in the medium surrounding the clusters of ionizing radiation sources with their characteristic properties and spatial distribution. The numerical procedures in turn are based on discretization schemes, the requisite accuracy of which determine the required lengths of the time steps, the extent of the cubic cells, the number and orientation of the rays, and the number and distribution of the frequency points. Using these iteration and discretization schemes the mean intensities are calculated per frequency interval for every cell by tracing all rays in sequential order, and from these intensities the occupation numbers are determined via a solution of the rate equations for each cell and time step. Thus, the implementation of the iteration cycle is governed by the geometrical aspects of the ray by ray solution and the time-dependencies of the equation schemes. To specify these items we will in the following section describe how the geometrical structure in our simulations is established in detail. In this regard special emphasis will be given to the ray distribution used for the radiative sources, and to the tedious calculation of the lengths of the ray segments within the cells (Sect. 3.1). The accuracy of our method will finally be tested in this section by some benchmark tests (cf. Sect. 3.3).

### 3.1. Ray tracing in a cartesian grid

The fundamental geometrical task in grid- and ray-tracing-based 3-dimensional radiative transfer is to distribute the rays correctly. The two main geometrical aspects to meet this task are (a) tracing the segments of each individual ray correctly (cf. Sect. 3.1.1) and (b) implementing the isotropic character of the radiation field of every point source by distributing the rays as evenly as possible (cf. Sect. 3.1.2), furthermore assuring that each cell is traversed by a sufficient number of rays[22] (cf. Sect. 3.1.3). With respect to the second concern we have applied and tested two different approaches: a conventional latitude/longitude method where the directional vectors are equally-spaced in polar angle, and for each polar angle equally-spaced in azimuth (cf. Abel et al. 1999), and a method based on dividing the surface of a sphere into elements of equal area (Górski et al. 2005) and using the directions of the centers of these areas as directional vectors for the rays. As these tests have shown that the latter method distributes the rays more evenly and thus leads to less discretization artifacts, we have decided to use this one.

---

[20] We note that the structure of the photoionization cross-sections demands the photoionization integrals (Eq. 20) to be evaluated on a logarithmic wavelength scale.

[21] The simulated volume, which is filled with hydrogen with a number density of 1 cm$^{-3}$ and a gas temperature of 10 000 K, has an edge length of 101 pc.

[22] Otherwise, a cell might not notice the presence of the radiation field (cf. Fig. 3).

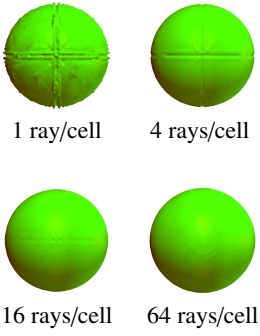

**Fig. 5.** Computed ionization front around a single source in a homogeneous medium as a function of increasingly fine discretization in angles. The volume consists of $61^3$ grid cells and is seen pole-on (i.e., from along the $z$ axis). Deviations from the ideal spherical shape are particularly obvious close to the planes defined by the coordinate axes in the models with a low number of rays (top row). Only with a sufficient number of rays can the models reproduce the physically expected spherical shape (bottom row).

1 ray/cell, 4 rays/cell, 16 rays/cell, 64 rays/cell

### 3.1.1. Tracing the segments of a single ray

A source at coordinates $(x_S, y_S, z_S)$ embedded in a cartesian grid of cubic cells of identical size emits rays whose directions are defined by $\vartheta$, the angle between the directional vector and the "north pole" (positive $z$ axis), and the azimuthal angle $\varphi$ (measured from the $x$ axis). To compute the radiative transfer along the rays we have to know which cells are passed by which rays and how long is the segment of a ray that crosses a cell. To obtain this information we follow each ray from its source to the edge of the simulation volume and determine in sequence which border between grid cells will be crossed next by the ray. This is done via a calculation of the distances $r_{yz}$, $r_{xz}$, $r_{xy}$ to the next $yz$-, $xz$-, $xy$-plane crossings in the direction of the ray. (For this purpose it is convenient to choose a coordinate system where the unit is equal to the length of a cube; in the simulation the actual physical length of a grid cell is then given by a simple scaling law.) The values $r_{yz}$, $r_{xz}$, $r_{xy}$ thus indicate the distances from the origin of the source to the next point where a grid cell is crossed and the shortest of these distances is the one which defines the ray segment $l(m)$ (cf. Eq. 4 and Eq. 25).[23]

### 3.1.2. Distribution of the rays

As the radiation field of an ideal point source is isotropic, any discretization of the radiation field should approximate this as close as possible and distribute the rays as uniformly as possible. Furthermore, as the luminosity of the source has to be divided among all rays, it would be highly advantageous if all rays were to correspond to approximately the same solid angle as seen from the source. Thus we have selected a method where the surface of a sphere around the radiation source is split into almost uniformly distributed segments of equal area, so that the directional vectors to the segment centers correspond to equal solid angles and are almost isotropically distributed ("hierarchical equal-area iso-latitude pixelization" method, HEALPix; Górski et al. 2005). These vectors are then used as the directions of the rays.[24]

### 3.1.3. Exploitation of symmetries

One of our modifications of the algorithm outlined above for computing the geometry of the rays regards the symmetry of the ray distribution of the individual sources. Because of this symmetry it is sufficient to compute the directional vectors just in one octant and to mirror the calculated geometrical factors at the planes defined by the coordinate axes.[25] By choosing the size of the octant as large as the entire simulated volume, the same geometrical information can be used for all sources assumed within the volume.

Besides the distribution of the rays the actual number of rays used is essential for the quality of the radiative transfer, because higher ray densities lead to a better representation of isotropy with less discretization artifacts, and in the (theoretical) limit of an infinite number of rays, the radiation field defined by the rays correctly describes the true (continuous) radiation field. Contrariwise, isotropy can be badly violated by using an insufficient number of rays, even if a good distribution algorithm is chosen. The dependence of the accuracy of the radiative transfer on the number of rays is demonstrated in Fig. 5.

For this test case we have chosen a single ionizing source embedded in a homogeneous isothermal gas. Because of symmetry we know that the resulting ionized volume in the gas must be spherical. (As the used geometry is independent of the density structure of the gas, the results regarding the minimum ray density necessary for an accurate description are of course also

---

[23] With the projections of the unit vector in the direction of the ray on the axes $e_x = \cos\varphi \sin\vartheta$, $e_y = \sin\varphi \sin\vartheta$, $e_z = \cos\vartheta$, and $\sigma_x = \text{sign}(e_x)$, $\sigma_y = \text{sign}(e_y)$, $\sigma_z = \text{sign}(e_z)$, one gets for the distances $r$ to the next plane crossings (with $r_{yz|xz}$ both, $r_{yz}$ and $r_{xz}$ are identified)

$$r_{yz|xz}(m+1) = \frac{\left|(p_{x|y}(m) + (\sigma_{x|y}+1)/2) - x_S|y_S\right|}{\max(\varepsilon, |\cos\varphi \sin\vartheta|)},$$

$$r_{xy}(m+1) = \frac{\left|(p_z(m) + (\sigma_z+1)/2) - z_S\right|}{\max(\varepsilon, |\cos\vartheta|)},$$

where $m$ is a counter variable for the cells that have been crossed by the ray, and the $\varepsilon$-terms are introduced to prevent a division by zero (cf. Abel et al. 1999; $\varepsilon$ is chosen to be a small positive number), and the position indices of the crossed cells relative to the cell of the source are indicated by the components $p_x, p_y, p_z$. Along with $s(m+1)$, the length of the ray at the end of the current cell (cf. Eq. 4 and Eq. 6), the new position indices $p_x, p_y, p_z$ (of the exit point of the ray ≜ entrance point to the next cell) are obtained from:

- if $r_{yz|xz|xy}(m+1) \leq \min(r_{xz|yz|yz}(m+1), r_{xy|xy|xz}(m+1))$, then
  $$s(m+1) = r_{yz|xz|xy}(m+1) \qquad p_{y|x|x}(m+1) = p_{y|x|x}(m)$$
  $$p_{x|y|z}(m+1) = p_{x|y|z}(m) + \sigma_{x|y|z} \qquad p_{z|z|y}(m+1) = p_{z|z|y}(m);$$

and one gets for the ray segments used in Eq. 22 $l(m) = s(m+1) - s(m)$. This procedure is repeated for each ray $n$. Finally, all ray segments are scaled by the corresponding physical size of the unit cell.

[24] By dividing the sphere into two polar caps and an equatorial region the rays lie here on $4N_R - 1$ iso-latitude rings, and each of the $2N_R + 1$ rings in the equatorial zone contains a constant $4N_R$ rays, whereas the number of rays per latitude in the polar regions increases with increasing distance from the pole by 4 per ring (controlled by a resolution parameter $N_R = 1, 2, 4, 8, \ldots$ the total number of rays therefore is $N = 12N_R^2$). The directions of the rays, $\vartheta$ measured from the north pole (positive $z$ axis) and $\varphi$ measured from the $x$ axis, are then given by

$$\cos\vartheta = 1 - \frac{1}{3}\left(\frac{i}{N_R}\right)^2 [i=1,...,N_R-1], \varphi = 2\pi \frac{j-\frac{1}{2}}{4i} [j=1,....,4i],$$

in the north polar cap ($2/3 < \cos\vartheta \leq 1$), and
with $\xi(i) = ((i - N_R + 1) \mod 2)$ by

$$\cos\vartheta = \frac{2}{3}\left(2 - \frac{i}{N_R}\right) [i=N_R,...,2N_R], \varphi = 2\pi \frac{j-\frac{\xi(i)}{2}}{4N_R} [j=1,...,4N_R].$$

in the northern equatorial region ($0 \leq \cos\vartheta \leq 2/3$) including the equator. These are mirrored in $z$ for the rays in the southern hemisphere.

[25] We have also tested a method in which segments of multiple rays are merged near the source to reduce the amount of ray segments that need to be processed (hierarchically ordered splitting procedure as described by Abel & Wandelt 2002). However, we found this description for the radiation field less accurate than the one without ray splitting/merging, and the additional administrative overhead not worth the small reduction in computing time achieved.

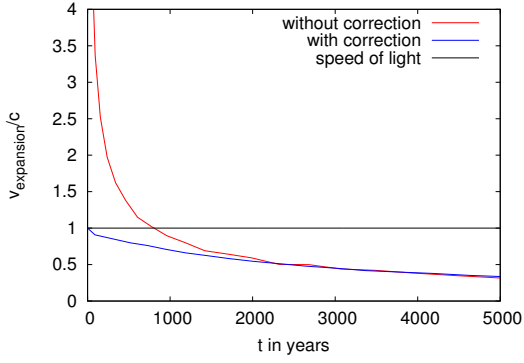

**Fig. 6.** Comparison of the corrected expansion velocity of the ionization fronts (blue line) with the uncorrected one (red line) in units of the speed of light. The correction is necessary, because the explicitly time dependent term appearing in the equation of radiative transfer is not considered in our simulations. As a consequence the ionization fronts can in an artificial way propagate faster than the speed of light (cf. Sect. 3.2). These quasi-super-luminous expansions of the ionization fronts are avoided by delaying to the correct time step the computation of the radiative transfer for cells that can not in physical reality be reached by light emitted by the source since its appearance. For the setup of the simulation a black-body radiator with an effective temperature of 95 000 K and a radius of 9 solar radii embedded in a pure hydrogen gas with a particle density of $n = 10^{-4} \text{cm}^{-3}$ and a temperature of 10 000 K has been assumed.

applicable for non-symmetric problems as well.) We have found that the deviations from the ideal sphere are maximal along the planes defined by coordinate axes. This is the case for both the HEALPix method as well as the equal polar angle intervals method. For an accurate representation (within the limits of the cartesian cell grid) of the true shape of the ionized volume we need at least 10 rays per cell even in the corners of the simulation volume farthest removed from the source.

### 3.2. Accounting for the finite speed of light

Although the geometrical properties of the rays are essential for a correct description of the radiative transfer, the evolution of the ionization structures are primarily determined by the ionizing luminosities of the sources; and this means that for conditions where the ratio of photon emission rate to the gas density is very high the radius of the physical ionization front can propagate even at velocities which are close to the speed of light $c$. For cases where recombinations are extremely rare (e.g., at low gas densities) this behavior can even analytically be described: As essentially every emitted photon ionizes one hydrogen atom, which under these idealized conditions then remains permanently ionized, the volume of the ionized region, $V = (4\pi/3)r^3(t)$ as a function of time, multiplied by the particle number density, $n$, will be equal to the total number of emitted photons, $\dot{N} \cdot t$. Thus, assuming spherical symmetry

$$\dot{N} \cdot t = V \cdot n = \frac{4\pi}{3} r^3(t) \cdot n. \quad (28)$$

The time-dependent radius of the ionization front and its temporal derivative therefore are

$$r(t) = \sqrt[3]{\frac{3\dot{N}t}{4\pi n}} \quad , \quad \frac{dr}{dt} = \sqrt[3]{\frac{3\dot{N}}{4\pi n}} \frac{d}{dt} t^{1/3} = \frac{1}{3} \sqrt[3]{\frac{3\dot{N}}{4\pi n}} t^{-2/3}. \quad (29)$$

But, at the beginning of the irradiation this simple model predicts an expansion rate of the ionization front which is faster than the

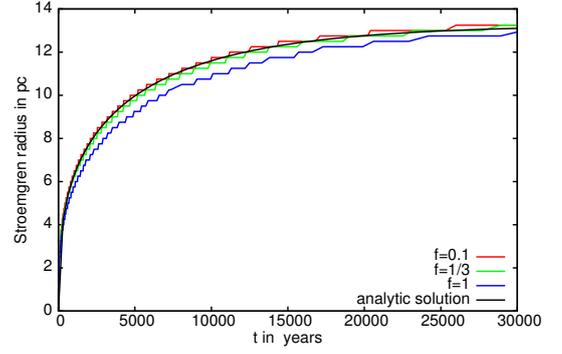

**Fig. 7.** Time-dependent evolution of a Strömgren sphere in a hydrogen gas ($n = 10 \text{ cm}^{-3}$, $T = 7\,500$ K) surrounding a hot star described by a blackbody radiator with an effective temperature of 40 000 K and a radius of 9 solar radii for different values of the parameter $f$ which controls the time step size (cf. Eq. 15). As is shown, the simulated expansion of the ionization front becomes slower for increasing time steps, and the analytic solution for the expansion of the ionization front described by Eq. 31 (black line) is for a value of $f = 0.1$ quite well represented by our simulation (red curve).

speed of light:

$$\frac{dr}{dt} > c \text{ for } t < \frac{1}{3}\sqrt{\frac{\dot{N}}{4\pi n c^3}}. \quad (30)$$

Since physical ionization fronts can not really propagate faster than the speed of light, we avoid such quasi-super-luminous expansions of the ionization fronts by delaying to the correct time step the computation of the radiative transfer for cells that can not in physical reality be reached by light emitted by the source since its appearance. This method is reasonable if, after being turned on, the luminosities of the sources do not vary considerably during the light travel time through the ionized volume. Figure 6 shows the speed of the ionization front as a function of its radius computed using our method, taking into account both recombination as well as the above described correction procedure.

### 3.3. Benchmark Tests

To test the accuracy of our method we have performed a series of calculations which we compare in this section to an analytical approximation, results from other 3-dimensional radiative transfer codes published in the literature (Iliev et al. 2006), and a detailed spherically symmetric model.

#### 3.3.1. Temporal expansion of a Strömgren sphere in a homogeneous hydrogen gas

A very simple test case is offered by the expansion of a Strömgren sphere in a homogeneous isothermal medium consisting of pure hydrogen, since under these approximations the temporal behavior of the sphere can be described analytically

$$r(t) = \sqrt[3]{\frac{3\dot{N}\left(1 - e^{-n\alpha_B t}\right)}{4\pi n_H^2 \alpha_B}}, \quad (31)$$

where $\dot{N}$ is the number of emitted ionizing photons per time unit, $\alpha_B$ is the case-B recombination coefficient and $n_H$ is the number density of hydrogen (Iliev et al. 2006). In Fig. 7 we compare this analytical solution with the result of a series of our simulations produced by a variation of the parameter $f$ which controls the time step size (cf. Eq. 15). The comparison shows on the one hand that for $t \to \infty$ the asymptotic value $r(t)$ is almost identical with the analytic solution, on the other hand it is also shown that the usage of larger time steps leads to slower expansions of the simulated Strömgren spheres. The explanation of this behavior is based on the fact that while the ionization fractions are updated during a time step the radiative transfer is not, and reflects the (higher) opacities at the beginning of the time step. Therefore, the number of photons reaching the cells just behind the ionization front is underestimated for long time steps. (Still, very large time steps can be used if the aim of the simulation is to compute the equilibrium state of the irradiated gas instead of an accurate description of the temporal evolution of the ionization structure.)

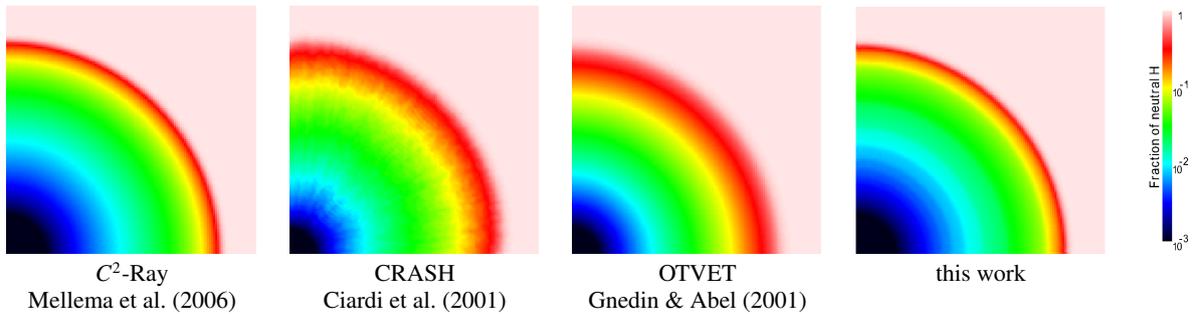

**Fig. 8.** Ionization structure in a pure hydrogen gas from four model codes using different approaches for computing the radiative transfer: $C^2$-Ray and our method (ray-tracing), CRASH (Monte Carlo), and OTVET (variable Eddington tensor method). The model parameters are as specified by Iliev et al. (2006): particle density $10^{-3}$ cm$^{-3}$; illumination by a monochromatic source emitting $5 \cdot 10^{48}$ hydrogen-ionizing photons/s; box size 6.6 kpc. Both ray tracing codes give practically identical results.

### 3.3.2. Comparison with other 3-dimensional radiative transfer methods

To compare the ionization structure from our code with that from other 3d radiative transfer codes we have computed models using the parameters specified in the cosmological radiative transfer comparison project (Iliev et al. 2006). Fig. 8 shows our result for the ionization state of a pure hydrogen gas enclosed in a box of the size of 6.6 kpc and illuminated by a monochromatic ionizing source in comparison with other simulations, namely the $C^2$-Ray simulation, which also uses the ray-tracing method, the CRASH simulation, which utilizes a Monte Carlo technique, and the OTVET simulation, which uses the variable Eddington tensor method. Although the results of the various methods do not reveal identical shapes, the fact that the structures of the $C^2$-Ray simulation on the left hand side of the figure and ours on the right hand side of the figure are almost indistinguishable is a mutually satisfactory result, since both procedures are based on the same technique. (Other codes based on ray tracing give similar results to these two.)

### 3.3.3. Comparison of cartesian and spherically symmetric models

For a test which also includes helium as a gas component in our 3d radiative transfer model we used a spherical symmetric equilibrium model (cf. Hoffmann et al. 2012) as reference. We present two sets of models, one using a hydrogen number density of $10$ cm$^{-3}$ (a typical value for H II regions) and one using a hydrogen number density of $10^{-3}$ cm$^{-3}$ (a typical value for a cosmological density field at early epochs). (As usual we set the He number fraction to $n_{He}/n_H = 0.1$.) For both gas densities we show in Fig. 9 resulting ionization structures in the gas: in the upper panels for a $T = 30\,000$ K blackbody emitting $10^{49}$ hydrogen-ionizing photons per second, representing a late O-type star, and in the lower panels for a 65 000 K blackbody emitting $8 \cdot 10^{51}$ hydrogen-ionizing photons per second, representing a very massive star (cf. Sect. 4.1).

In those models where we make use of the on-the-spot approximation in the spherically symmetric models, we get exactly the same results – within the resolution of the grid – as for our 3d calculations: not only are the Strömgren radii the same, but also the radius-dependent ionization fractions of H and He are almost identical for both the soft and the hard spectrum source. If we don't apply the on-the-spot approximation and instead treat the diffuse radiation field correctly in the spherically symmetric models, we find deviations on the order of 20 % for the occupation numbers of the neutral stages of hydrogen and helium in the inner regions. The reason for this is that part of the reemitted photons is not reabsorbed locally, as suggested by the on-the-spot approximation, but, due to the low number of available absorbers, at larger distances. However, these small deviations are irrelevant with regard to the positions of the ionization fronts and thus the sizes of the Strömgren spheres.

## 4. Photoionization models for H II regions around Luminous Stars in the Early Universe

With this new method in hand we will present in the following first exemplary simulations of our 3d radiative transfer applied to the reionization scenario. As different characteristics of the spectral energy distributions of the sources of ionization in general have a considerable impact on the power of ionization, the application of oversimplified spectra in radiative transfer codes – e.g. black-body radiators which are based on just a few discrete wavelengths – can lead to incorrect or deceptive results. In order to quantify such possible systematic errors in the context of hot massive stars we first will perform a comparison of blackbody and realistic spectral energy distributions at different metallicities stressing for the ionizing spectra of the relevant stellar populations used as ionizing sources the importance of metal enrichment. As our focus lies on a high spectral resolution of the 3d radiative transfer describing the evolution of the most important ionization structures of the IGM in certain epochs, we will in this concern not just concentrate on a massive star representing the top of a Salpeter IMF, but also on a very massive star (VMS) resulting from a cluster-collapse as a runaway collision merger (cf. Pauldrach et al. 2012 and references therein).

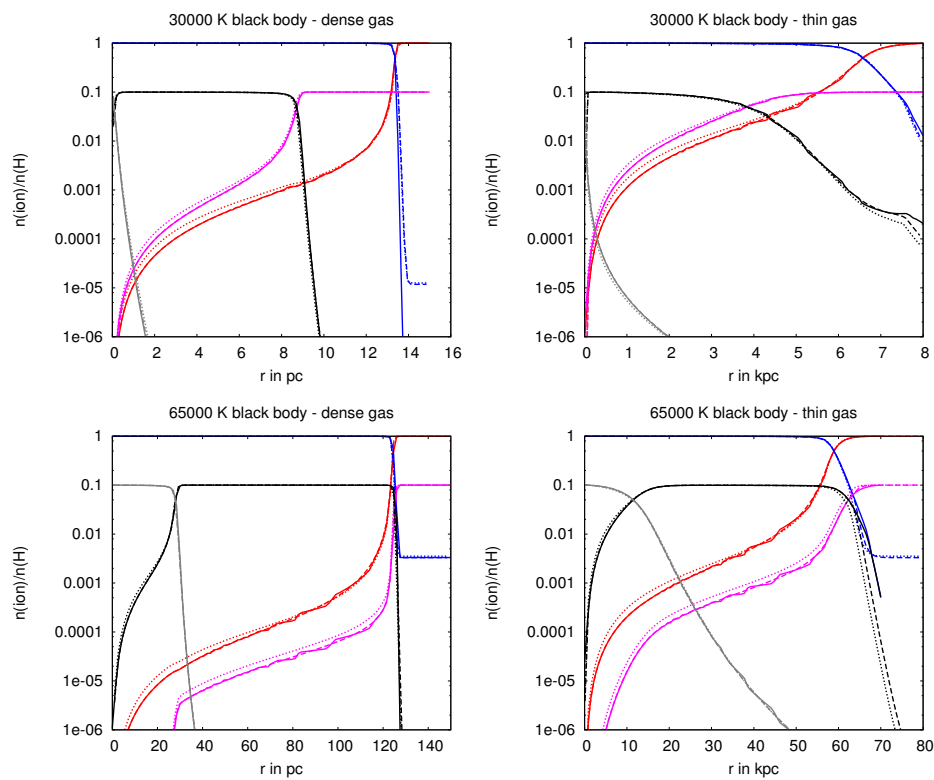

**Fig. 9.** Strömgren radii and the radius-dependent H and He ionization fractions calculated with our 3d radiative transfer method (solid lines) compared to those from our spherically symmetric method. For the latter we use either the on-the-spot approximation (dashed lines) or a correct description of the diffuse radiation field (dotted lines). In the upper panels the ionizing source is a $T = 30\,000$ K blackbody emitting $10^{49}$ hydrogen-ionizing photons per second, representing a late O-type star (the temperature of the gas has been fixed to $7\,500$ K), and in the lower panels a $65\,000$ K blackbody emitting $8 \cdot 10^{51}$ hydrogen-ionizing photons per second, representing a VMS in the lower panel (the temperature of the gas has been fixed to a value of $10\,000$ K). The gas densities are $n_\mathrm{H} = 10\,\mathrm{cm}^{-3}$ for the plots on the left and $n_\mathrm{H} = 10^{-3}\,\mathrm{cm}^{-3}$ for those on the right.

In the introduction we discussed that the reionization process may have occurred in two distinct steps (Cen 2003), where the first one was primarily driven by population III stars and was concluded at a redshift of $z \sim 15$, whereas the second one, which completed the reionization with radiation from population II/I stars, finished at a redshift of $z \sim 6$. With respect to time-dependent simulations of the reionization scenario we will therefore perform in a first step multi-source simulations for population III stars in homogeneous and inhomogeneous cosmological gas distributions representing the first stage of the reionization process in order to investigate on the one hand the evolution of the ionization fronts into the IGM and on the other hand the fraction of photons which can escape from the ISM surrounding the ionizing sources and thus build up the ionization fronts in the IGM. In an advanced step we further will combine our predicted SEDs of massive and very massive stars with time-dependent 3d radiative transfer simulations of the IGM in order to describe the second stage of reionization. Although our models are still based on simplified initial conditions, the general requirements that might have allowed the much less massive population II and population I stars to fully reionize the IGM – this approach is basically motivated by the increasing star formation rate in this epoch (Lineweaver 2001) – can already be investigated even at this stage. Regarding the reionization of He II to He III in particular, we investigate further the influence and relevance of a top-heavy-IMF connected to very massive stars. Whether the appearance of population II and population I VMSs could have been responsible for the reionization of He II will therefore be examined as a crucial point.

### 4.1. Comparison of blackbody and realistic spectral energy distributions at different metallicities as ionizing sources

Motivated by the fact that metallicities which are different from zero can have a strong influence on the spectral energy distributions (SED) of massive stars and that the ionizing fluxes can at certain frequencies ranges therefore depart decisively from those of blackbody radiators we computed comparison models using synthetic SEDs from atmospheric models instead of blackbodies as ionizing sources. To limit the scope of this challenge we do not attempt to trace the metallicity dependent characteristics of the spectral energy distributions of a complete set of stellar populations we rather want to demonstrate the effects of metallicity on the spectra and furthermore the ionized structures by means of a series of striking objects.

A massive star representing the top of a Salpeter IMF. As striking stellar objects are primarily characterized by their mass, we have chosen a $125\,M_\odot$ star with an effective temperature of $T_\mathrm{eff} = 50\,000$ K and a radius of $R = 21.7\,R_\odot$ as a first example. Figure 10 shows spectral energy distributions computed for this object assuming different metallicity values (ranging from 0.001 to 1.0 $Z_\odot$ – the spectra have been calculated on basis of the method described by Pauldrach et al. 2001, 2012). With respect to the mass each of these models represents obviously a star which can be regarded to mark the top of a metallicity dependent Salpeter IMF. As suspected, Fig. 10 shows clearly that an enhancement of metallicity has a drastic influence on the SEDs. While the hydrogen-ionizing flux ($\lambda < 911$ Å) is primarily influenced by P-Cygni spectral lines, shorter wavelengths, as the He II ionization edge at $\sim 228$ Å, are also influenced by NLTE-effects affecting the bound-free thresholds of certain ionization stages (cf. Pauldrach et al. 2012) and producing in comparison with blackbody radiators considerable differences.

The results of our 3d radiative transfer calculations for the radial density distribution of ionized hydrogen surrounding the $125\,M_\odot$ stars with different metallicities are shown in Fig. 11. (The parameters of the model correspond to a typical H II region: $n_\mathrm{H} = 10\,\mathrm{cm}^{-3}$, $n_\mathrm{He}/n_\mathrm{H} = 0.1$, and $T_\mathrm{gas} = 10\,000$ K.) Although at the effective temperature of $50\,000$ K of the ionizing source the difference of the solar metallicity spectrum to the blackbody spectrum leads to a considerable difference of ∼ 4% for the hydrogen Strömgren radius – corresponding to a difference of ∼ 12% in the ionized volume –, the variation of the metallicity of the stars does not have a huge influence on the number of hydrogen-ionizing photons, and therefore a replacement of especially the lowest metallicity spectrum (0.001$Z_\odot$) by a blackbody spectrum, which leads to a radial error of ∼ 3% which corresponds to a maximal error of ∼ 9% in the ionized volume, may appear to be legitimate.

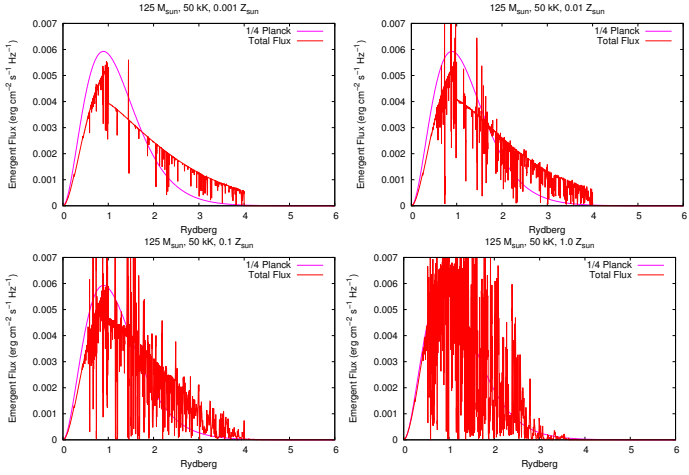

**Fig. 10.** Emergent Eddington flux $H_\nu$ versus wavelength calculated for 125 $M_\odot$ stars with effective temperature 50 000 K and different metallicities – 0.001, 0.01, 0.1 and 1.0 $Z_\odot$ (from upper left to lower right) – compared to an equivalent blackbody.

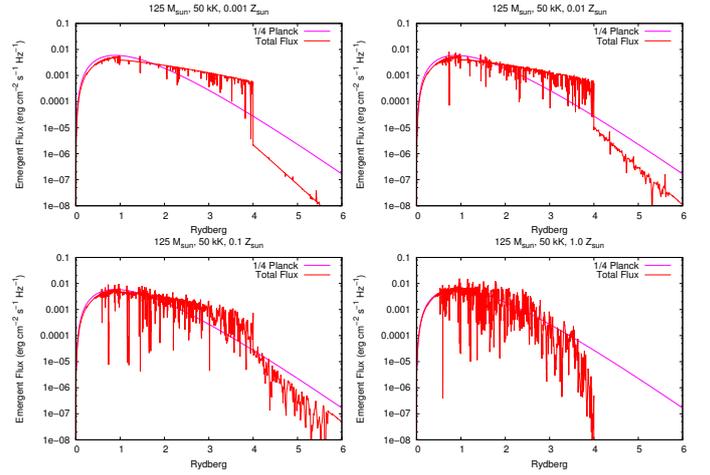

**Fig. 12.** As Fig. 10, but the flux is now shown on a logarithmic scale. On this scale it is verified that the ionizing fluxes shortward of the He II ionization edge (4 Rydberg) vary considerably along with a change of the metallicity (cf. Table 2).

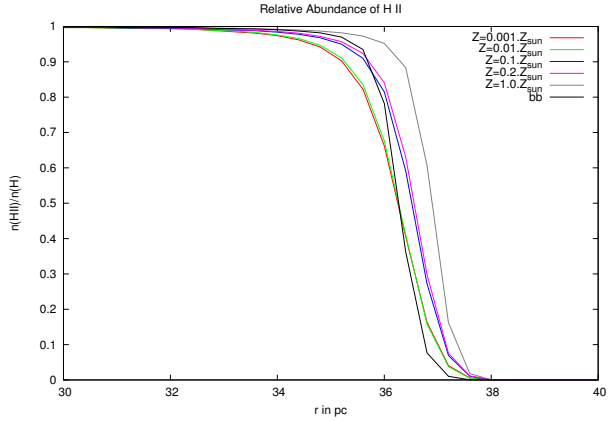

**Fig. 11.** Ionized hydrogen fraction in a H/He gas ($n_H = 10\,\text{cm}^{-3}$, $n_{He}/n_H = 0.1$, and $T_{gas} = 10\,000$ K) surrounding the above 125 $M_\odot$ stars, again in comparison to that of an equivalent blackbody. The variation of the stellar metallicity only has a small influence on the Strömgren radii.

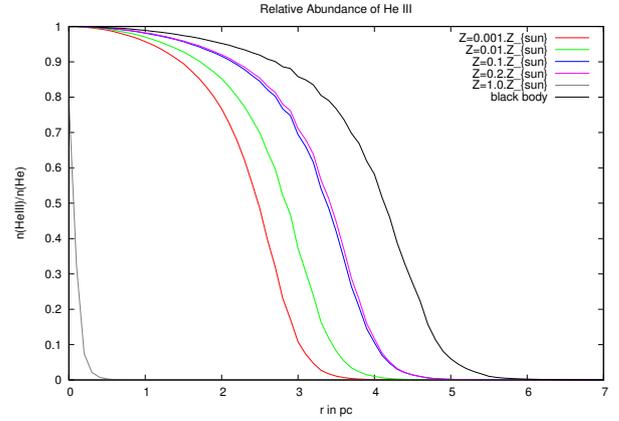

**Fig. 13.** As Fig. 11, but the ionization fraction of He III is now shown (cf. Table 2). In contrast to hydrogen the efficiency to ionize He II depends in this stellar parameter range crucially on the metallicity of the stars, leading to a strong influence on the Strömgren radii.

**Table 2.** Logarithm of the number $Q$ of H-, He I-, and He II-ionizing photons emitted per second by a massive star ($M = 125\,M_\odot$, $R = 21.7\,R_\odot$, and $T_{eff} = 50\,000$ K) as a function of metallicity $Z$, compared to those of an equivalent black body. The third column lists the consistently calculated mass-loss rates $\dot{M}$ of the stellar models.

| $Z/Z_\odot$ | $\dot{M}$ ($10^{-6}\,M_\odot$/yr) | $\log Q_H$ | $\log Q_{He\,I}$ | $\log Q_{He\,II}$ |
|---|---|---|---|---|
| 1.0 | 29.0 | 50.28 | 49.66 | 42.84 |
| 0.2 | 13.3 | 50.26 | 49.77 | 46.90 |
| 0.1 | 9.2 | 50.26 | 49.78 | 46.89 |
| 0.01 | 2.0 | 50.25 | 49.81 | 46.56 |
| 0.001 | 0.1 | 50.25 | 49.82 | 45.91 |
| black body |  | 50.25 | 49.54 | 47.14 |

However, the similarity of the hydrogen Strömgren spheres presented here results primarily from the fact that the influence of the metals in the gas and therefore the corresponding cooling and heating processes have been neglected. A consistent computation including metals would actually lead to smaller temperatures for larger metallicities and thus result in smaller Strömgren spheres (Osterbrock & Ferland 2006). An even larger influence on the ionization structure of the metals in the gas and thus on the gas temperatures and Strömgren spheres will be caused by the non-LTE line blocking and blanketing[26] in the expanding at-mospheres of the massive stars which ultimately determine the shapes of the SEDs (cf. Sellmaier et al. 1996 and Pauldrach et al. 2001 – a superficial impression of this behavior can also be obtained by a thorough inspection of Fig. 10). From these arguments it is quite obvious that our Fig. 11 does not yet represent an ultimate result!

---

[26] Non-LTE refers to detailed modelling of the statistical equilibrium of the occupation numbers of the atomic levels, obtained from actually solving the system of rate equations, without assuming the approximation of local thermodynamic equilibrium (LTE). The effect of line blocking refers to an attenuation of the radiative flux in the EUV and UV spectral ranges due to the combined opacity of a huge number of Doppler-shifted metal lines present in massive stars in these frequency ranges. It significantly influences the ionization and excitation and the

**Table 3.** As Table 2, but for a very massive star ($M = 3000\,M_\odot$, $R = 79\,R_\odot$, and $T_{\text{eff}} = 65\,000$ K) as a function of hydrogen mass fraction $X$ and metallicity $Z$.

| $X$ | $Z/Z_\odot$ | $\dot{M}$ $(10^{-6}\,M_\odot/\text{yr})$ | $\log Q_{\text{H}}$ | $\log Q_{\text{He\,I}}$ | $\log Q_{\text{He\,II}}$ |
|---|---|---|---|---|---|
| 0.70 | 0.05 | 132. | 51.89 | 51.55 | 49.77 |
| 0.70 | 1.0 | 385. | 51.92 | 51.50 | 49.60 |
| 0.40 | 1.0 | 260. | 51.88 | 51.53 | 49.33 |
| black body | | | 51.89 | 51.39 | 49.66 |

Regarding the radial density distribution of He III differences of the Strömgren radii which originate from different metallicities of the ionizing sources are considerably more pronounced (cf. Fig. 13) – as an example, the difference of the blackbody to the spectrum calculated with solar metallicity leads to a difference in the He III Strömgren radius by a factor of 30. But these differences are not necessarily a monotonic function of increasing metallicity of the ionizing sources. With respect to our calculations the number of emitted He II-ionizing photons increases from the model with $Z = 10^{-3}\,Z_\odot$ to the model with $Z = 0.2\,Z_\odot$, whereas it decreases drastically – by 4 orders of magnitude – for the model with solar metallicity (cf. Figure 12 and Table 2, which show that there is no strict relationship between the metallicity and the emission of He II-ionizing photons). These results show clearly that the number of He II-ionizing photons is not just strongly influenced by the effective temperature and hence the mass of the relevant objects, but also by the evolution of the metallicity. Employing blackbody spectra in general therefore leads for the cases of helium and metals whose ionization energies exceed those of H I and He I to inaccurate ionization structures. As exactly the elements which are affected by this approximation are producing the characteristic and in principle observable emission lines, such a weak-point in the computation of the ionization structures would make as a consequence a comparison to observations extremely disputable (cf. Rubin et al. 1991, Sellmaier et al. 1996, and Rubin et al. 2007).

**A very massive star as a runaway collision merger from a cluster-core collapse.** As a second example of our investigation of the influence of realistic spectral energy distributions of massive stars on the ionization structures of the surrounding gas we have chosen an even more impressive stellar object: a $3000\,M_\odot$ star. Such stars are expected to form as runaway collision mergers in cluster-core collapses, and via feedback effects due to their enormous radiation and strong stellar winds they naturally have a considerable impact on their environment and the local star formation (cf. Sect. 1).

Figs. 14 and 16 show the SEDs of three model atmospheres of such a star ($L = 10^8\,L_\odot$, $T_{\text{eff}} = 65\,000$ K; cf. Pauldrach et al.

2012) computed for different metallicities ($Z = 0.05\,Z_\odot$ and $Z = 1.0\,Z_\odot$) and different hydrogen mass fractions $X$ (70 % and 40 %). Together with the consistently calculated mass loss rates the ionizing fluxes of the model stars are listed in Table 3. The luminosities of these VMSs exceed those of massive O stars at the top of a Salpeter or Kroupa-IMF by one to two orders of magnitude.

Although the hydrogen-ionizing fluxes of these models are close to those of a corresponding blackbody (Fig. 14 and Table 3), increasing metallicity and/or increasing helium mass fraction leads to a clear decrease of the He II-ionizing fluxes: while the metal-poor star with a normal helium content has a He II-ionizing flux 29 % larger than that of a corresponding blackbody, for the helium-enriched star with solar metallicity it is 53 % *smaller* (Fig. 16 and Table 3). As a consequence of this behavior the hydrogen Strömgren radii differ from the blackbody Strömgren radius by only 3 % (Fig. 15), but the He III Strömgren radii show differences to the corresponding blackbody value of up to 25 % (Fig. 17).

The differences between our very massive stars and a black body are small, however, when compared to the differences between our very massive stars and our normal $125\,M_\odot$ massive stars. The hydrogen-ionizing fluxes of the $3000\,M_\odot$ stars are a factor of 40 larger than those of the $125\,M_\odot$ stars, while the He II-ionizing fluxes of the $3000\,M_\odot$ stars exceed those of the $125\,M_\odot$ stars by a factor of 800 to several $10^6$, depending on the metallicity (Table 2 and Table 3). Due to the large luminosities and effective temperatures of the very massive stars the radii of the hydrogen Strömgren spheres thus differ by a factor of 3.5 and those of the He III Strömgren spheres by a factor of up to 100 (Fig. 15 and Fig. 17).

We conclude from this that the transition from normal massive stars to very massive stars could strongly influence the relative volumes of He III to H II. This assessment does not change if the fluxes are normalized to the same hydrogen-ionizing photon rate: in this case the He II-ionizing fluxes of the VMSs are still at least a factor of 20 larger. As this finding may be of importance for a possible scenario of the cosmic He II-reionization, we will return to this point in Sect. 4.2.2.

## 4.2. Time-dependent simulations for different characteristics of the reionization scenario

In order to better understand the reionization process of the universe, the evolution of the ionization fronts expanding in the IGM has to be examined in some detail using a numerical treatment of the radiative transfer and the microphysics of ionization and recombination. For this purpose we present in this section typical results of our 3d radiative transfer simulations for representative values of the densities and star formation rates. In these simulations we use clusters of massive population III stars (for the simulations representing the first stage of reionization) and population II/I stars (for the second stage of reionization) as ionization sources, surrounded by an environment reflecting the intergalactic cosmological gas.

### 4.2.1. Multi-source simulations for population III stars in homogeneous and inhomogeneous gas distributions

As already mentioned in the introduction, the time around $z \sim 13.5$ marks the transition from population III to population II/I stars and leads to a decline in the emitted radiative power. On the other hand, the star formation rate of population III stars was

---

momentum transfer of the radiation field through radiative absorption and scattering processes. As a consequence of line blocking, only a small fraction of the radiation is re-emitted and scattered in the outward direction, whereas most of the energy is radiated back to the surface of the star, leading there to an increase of the temperature ("backwarming"). Due to this increase of the temperature, more radiation is emitted at lower energies, an effect known as line blanketing.

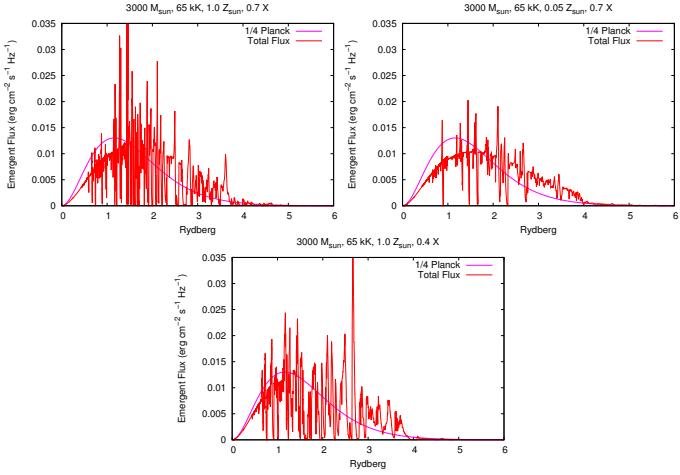

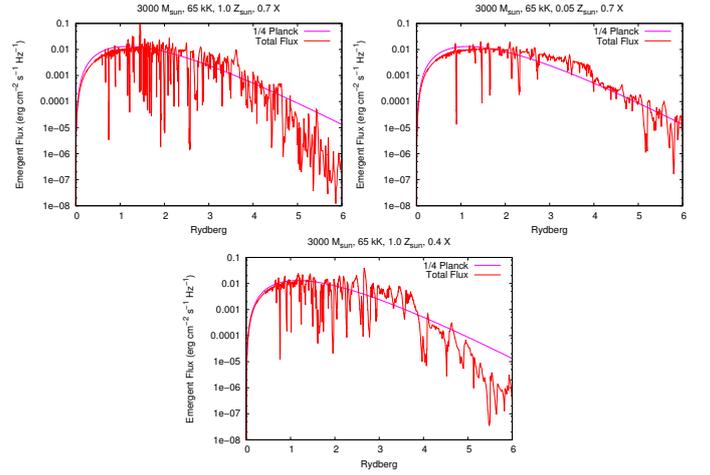

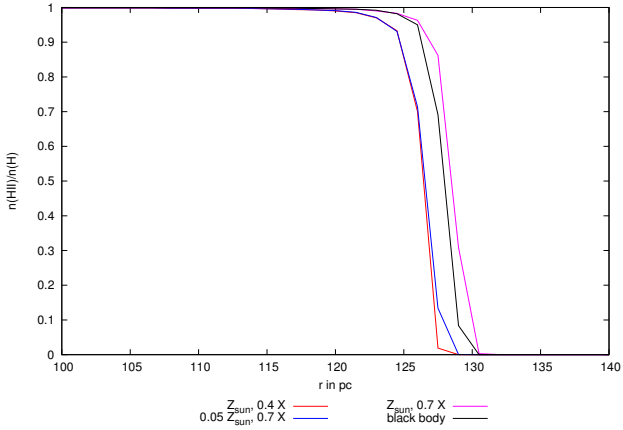

**Fig. 14.** Emergent Eddington flux $H_\nu$ versus wavelength calculated for very massive $3000\,M_\odot$ stars with effective temperature $65\,000$ K and radius $79\,R_\odot$, for different compositions (a metallicity of 1.0 and 0.05 $Z_\odot$ and a hydrogen mass fraction of 0.7 (first row), and a metallicity of 1.0 $Z_\odot$ and a hydrogen mass fraction of 0.4 (second row)), compared to an equivalent blackbody.

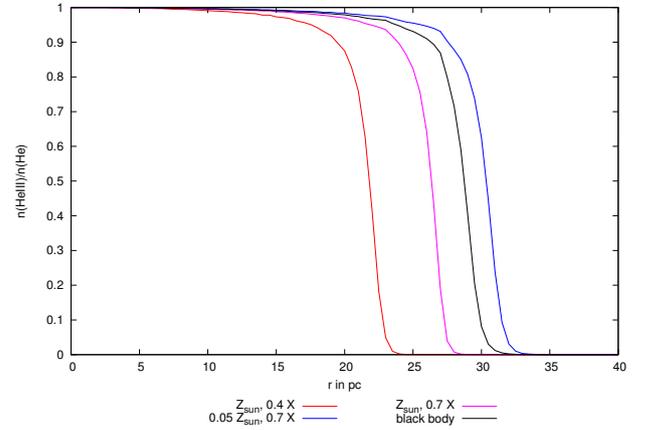

**Fig. 16.** As Figure 14, but the flux is now shown on a logarithmic scale. On this scale the variation of the fluxes in the range above the He II ionization edge with metallicity is verified. For a metallicity of $0.05\,Z_\odot$, there is a considerably stronger flux of these photons than for the models with $Z_\odot$. Because of the larger He abundance, the model at the bottom shows less emission in the He II-ionizing part of the spectrum.

**Fig. 15.** Ionized hydrogen fraction in a H/He gas ($n_H = 10\,\mathrm{cm}^{-3}$, $n_{He}/n_H = 0.1$, and $T_{gas} = 10\,000$ K) surrounding the above very massive stars, again in comparison to that of an equivalent blackbody. Here also, the variation of the stellar metallicity only has a small influence on the Strömgren radii.

**Fig. 17.** As Fig. 15, but the ionization fraction of He III is now shown (cf. Table 3). The efficiency to ionize He II rises for higher fractions of hydrogen and lower metallicities in the stars. This leads to Strömgren radii around VMSs which are a factor of up to 100 larger than those around "normal" $125\,M_\odot$ stars (cf. Fig. 13).

highest around $z \sim 15$, shortly before this stellar generation died out. For the first part of the reionization process that was driven by population III stars we have therefore selected this redshift for our calculations.

**First stage of the reionization scenario in a homogeneous gas distribution.** Representing the first stage of the reionization history we have calculated models which simulate a volume of $(330\,\mathrm{kpc})^3$ in proper coordinates at $z = 15$ (which is a volume of $(5.3\,\mathrm{Mpc})^3$ in comoving coordinates at $z = 0$) resolved into $101^3$ grid cells. The temperature of the gas has been kept constant ($T = 10\,000$ K, cf. Cen (2003)), and the simulations use randomly distributed clusters of massive population III stars surrounded by a primordial environment characterized – as a rough approximation of the real situation – by a homogeneous intergalactic cosmological gas with a hydrogen number density of $8 \cdot 10^{-4}\,\mathrm{cm}^{-3}$ (reflecting the mean baryonic density of the universe at the given redshift, cf. Yoshida et al. 2004). The spectra of the multiple ionizing sources with a representa-tive temperature of $95\,000$ K have been approximated by blackbodies (at such a high temperature and a negligible metallicity the number of hydrogen-ionizing photons is almost perfectly described by blackbody emitters, cf. Sect. 4.1). Although we have not accounted for structures on smaller scales in this simulation, we have compared the assumed distances of our clusters with the distances between the largest clusters investigated by Gnedin & Bertschinger (1996) and found reasonable agreement. Each of our ten clusters emits $4 \cdot 10^{53}$ hydrogen-ionizing photons per second, which corresponds to a star formation rate of $9.5 \cdot 10^{-3}\,M_\odot \mathrm{yr}^{-1}$ per comoving $\mathrm{Mpc}^3$ (matching the rate deduced from Cen 2003, and Hernquist & Springel 2003 at a redshift of $z = 15$) and represents the emission of $4\,000$ population III stars with a mass of $100\,M_\odot$ and a lifetime of $3$ Myr (El Eid et al. 1983, Schaerer 2002).[27]

---

[27] We note that our simulations are based on the assumption that the stellar population appears instantaneously and does not fade away. We further assume that there is an enhanced absorption of ionizing pho-

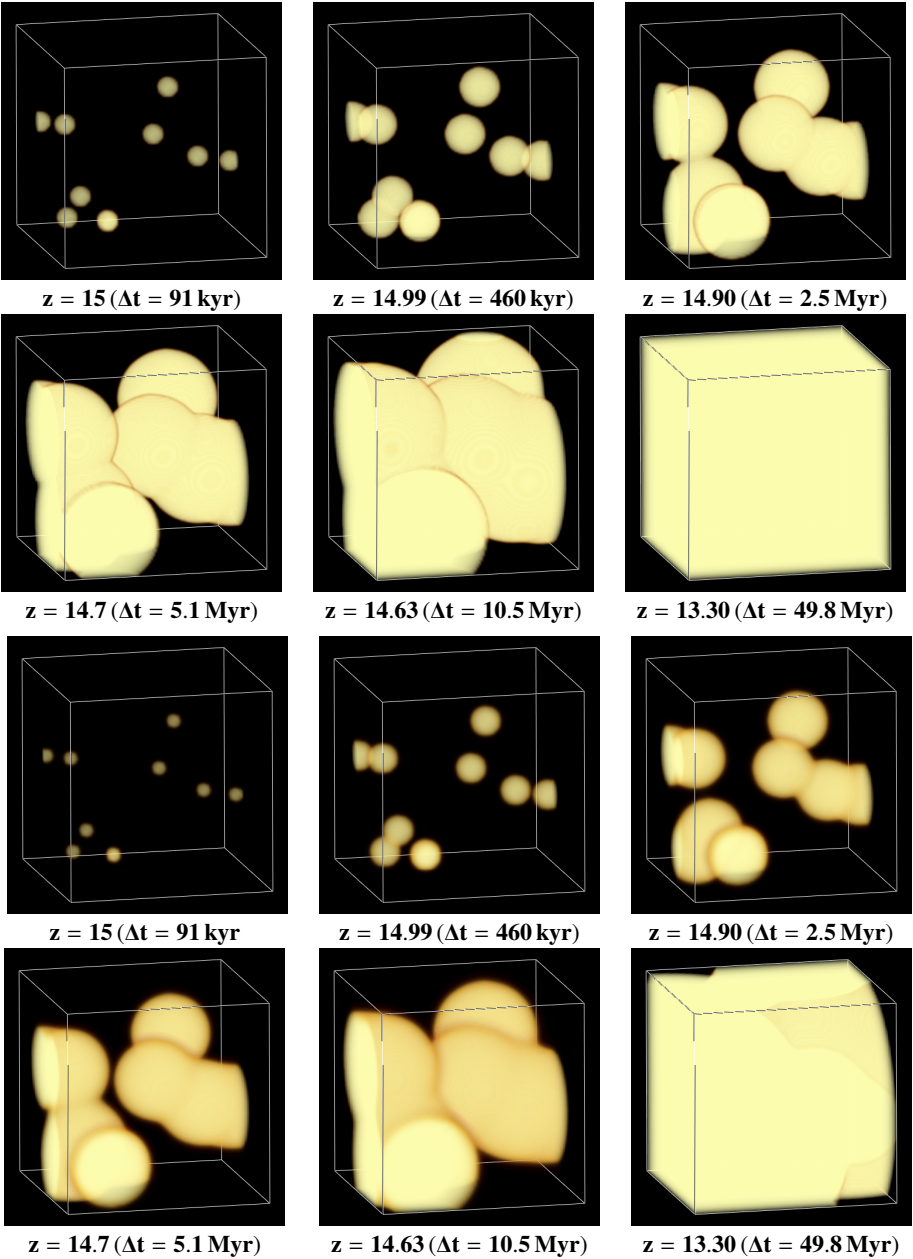

**Fig. 18.** The evolution of hydrogen ionization fronts expanding into the IGM in a simulation representing the first stage of the reionization history around $z = 15$. The volume is $(5.3\,\text{Mpc})^3$ in comoving coordinates, resolved in $101^3$ grid cells containing ten randomly distributed clusters of massive population III stars. Each cluster emits $4 \cdot 10^{53}$ hydrogen-ionizing photons per second, corresponding to a star formation rate of $9.5 \cdot 10^{-3}\,M_\odot\,\text{yr}^{-1}$ per comoving Mpc$^3$. The evolution of the ionization fronts starts with a set of discrete H II regions which begin to overlap after approximately 1 Myr. Hydrogen is completely ionized after 50 Myr (corresponding to $\Delta z \sim 1.5$ at $z = 15$).

**Fig. 19.** As Fig. 18, but showing He III. Due to the high effective temperatures (95 000 K) of population III stars and the correspondingly high emission rate of He II-ionizing photons the reionization of He II occurs on a timescale comparable to that of the reionization of hydrogen.

The evolution of the ionization fronts expanding into the IGM in our simulation is shown in Fig. 18. This simulation predicts that the clusters of population III stars as predicted by the theoretical star formation rate can ionize the intergalactic hydrogen on a timescale of just $t \sim 50$ Myr (corresponding to $\Delta z \sim 1.5$ at $z = 15$). Although cosmic variance could modify the reionization timescales when going to larger box sizes, this result is nevertheless significant, since our chosen volume with regard to the assumed and the observed mean distances of the cluster structures in this epoch is representative for the first stage of the reionization history of the universe. The time around $z \sim 15$ is the "optimum" period for this reionization phase since in this epoch the star formation rate of population III stars (Cen 2003), and thus the photon emission rate, rises rapidly, just before the changeover to population II stars significantly cuts down on the tons with respect to denser gas inside the clusters (from an investigation of smaller-scale structures surrounding the sources of ionization (cf. Fig. 22) we inferred that half of the ionizing photons should be absorbed within the denser gas inside the clusters for such an environment); thus, we assume that the escape fraction of the photons is 50 % ($f_{\text{esc}} = 0.5$).

emission rate of ionizing photons. Thus, our results indicate that the small contributions to the reionization before $z \sim 15$ are mostly insignificant for ionizing the universe completely, and the exact time of the beginning of the process is not essential.

As the large temperatures of massive population III stars result in considerably hard spectra, it is not surprising that the timescale of the reionization of He II into He III is not much different from the timescale of the reionization of hydrogen (each of our ten clusters emits $2 \cdot 10^{52}$ He II-ionizing photons per second, which is 1/20 of the amount of hydrogen-ionizing photons per second). Thus, if the first stage of reionization had really been driven, as currently suspected, by a stellar population dominated by massive hot primordial stars, helium would also have been reionized to a large extent into He III at a redshift of $z \sim 13$ (cf. Fig. 19). This would not have been the case, however, if the population III stars had a much softer spectrum (cf. Sect. 4.2.2) – as is indicated by some simulations which show a stronger fragmentation even for primordial gas (Clark et al. 2011). Whether helium had also been reionized into He III during the first stage – by a redshift of $z \sim 13$ – therefore depends crucially on the

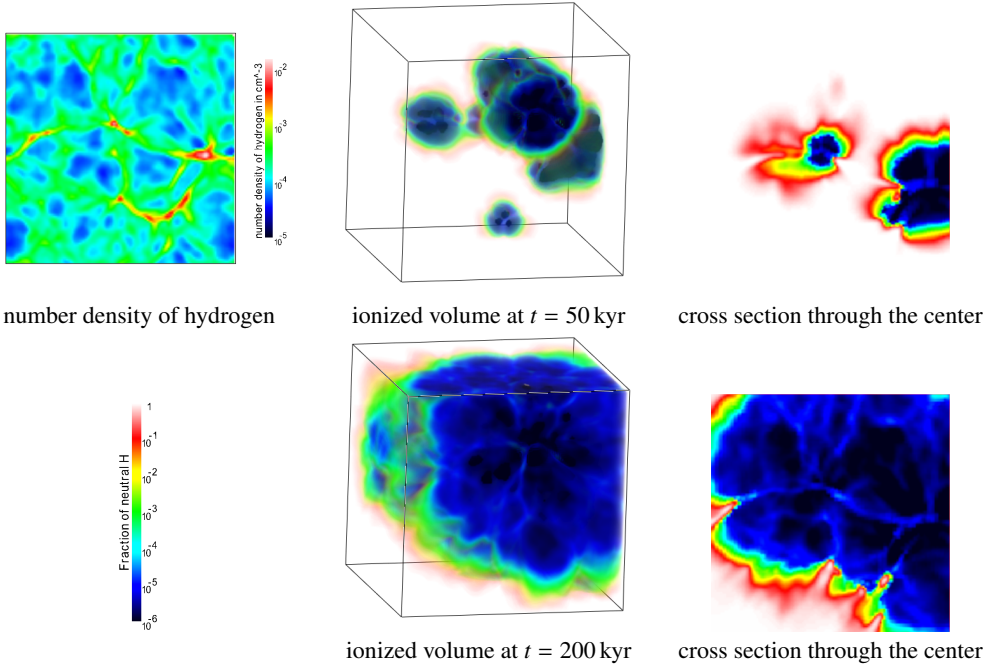

number density of hydrogen | ionized volume at $t = 50\,\mathrm{kyr}$ | cross section through the center

ionized volume at $t = 200\,\mathrm{kyr}$ | cross section through the center

**Fig. 20.** Simulation of the ionization of hydrogen in a volume of $(50\,\mathrm{kpc})^3$ (proper units) containing an inhomogeneous cosmological density field – the structure of the gas density has been obtained from simulations by Ryu et al. (1993). The image on the upper left hand side shows the number density of hydrogen along a plane through the center of the volume. The gas is illuminated by 16 clusters involving sources with an effective temperature of $95\,000\,\mathrm{K}$ and emitting a total of $\dot{N} = 4 \cdot 10^{53}$ hydrogen-ionizing photons per second. The ionization fraction of hydrogen is shown as 3d views of the ionized volume and as cuts along a plane through the center of the volume after $50\,\mathrm{kyr}$ and $200\,\mathrm{kyr}$.

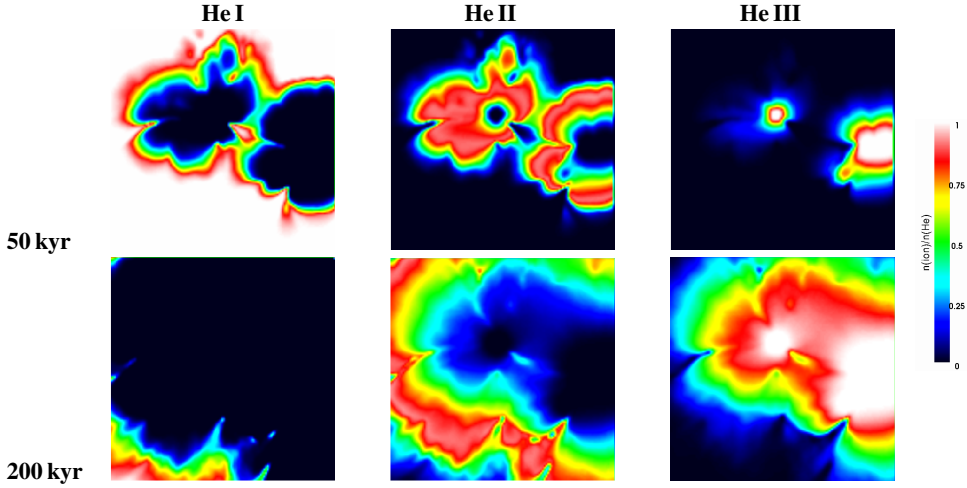

He I | He II | He III

50 kyr

200 kyr

**Fig. 21.** For the same simulation as in Fig. 20 cross-sections through the center, which display the relative abundances of the ionization stages of helium, are shown after $50\,\mathrm{kyr}$ and $200\,\mathrm{kyr}$.

emergence of massive population III stars in sufficient numbers. Whether these objects had been the dominant species of the stellar population that illuminated the very early universe can only be discovered by observations. In this regard the James Webb Space telescope as well as the European Extremely Large Telescope (E-ELT), the Giant Magellan Telescope (GMT) and the Thirty Meter Telescope (TMT) are expected to be able to observe clusters of bright population III stars at redshifts of $z \approx 13 \ldots 15$, and thus will help to improve our present knowledge of the stellar content governing the first stage of reionization.

Details of the reionization process in an inhomogeneous cosmological density field. It is obvious that the assumption of a homogeneous universe as applied to our models above is just a rough approximation of the real situation. As a consequence of this approximation effects which depend on small-scale inhomogeneities of the density structure of the gas and which in turn may considerably affect the reionization process can not be taken into account. Examples of such effects are the interrelation of the density structure and the local star formation rate – mutual back-reactions of the processes involved give rise to the inhomogeneous behavior of the ISM and are responsible for the presence of H II regions –, and the "porosity" of the medium which determines the escape fraction of ionizing photons that can exit their local environment and reach the intergalactic regions.

In order to study in particular the latter point, and thus the influence of multiple sources on their environment consisting of an inhomogeneous density structure on smaller scales, we have calculated a model which simulates in proper coordinates a volume of $(50\,\mathrm{kpc})^3$ resolved into $128^3$ grid cells. The density structure used for this simulation is based on results of hydrodynamical simulations from an $N$-body gas dynamics code (cf. Ryu et al. 1993 and Iliev et al. 2006). As this density profile had been calculated for cosmological conditions which are consistent with a redshift of $z = 9$ only, we rescaled it to conditions corresponding to a redshift of $z = 15$. The sources, which are located at the peaks of the structured density profile (pink dots in the upper left image of Fig. 20), cover 16 smaller clusters with a total emission rate of $\sim 4 \cdot 10^{53}$ hydrogen-ionizing photons per second (thus, all of the clusters in this simulation collectively represent one of the clusters in the homogeneous case (cf. Fig. 18)). As before, they are treated as blackbody radiators representing stars with an effective temperature of $95\,000\,\mathrm{K}$.

The results of this simulation, for which the gas temperature has again been held fixed at a value of $10\,000\,\mathrm{K}$, are pre-sented in Fig. 20 for hydrogen and Fig. 21 for helium. The overall impression of the images reflects the behavior of considerably more complex ionization structures for the inhomogeneous density structure compared to those of the homogeneous models. This impression is certainly true and the irregular form of the ionization fronts and the larger fraction of neutral hydrogen in the higher-density filaments is evident. The simulations also show that the ionization stages He II and He III coexist over a considerable part of the volume (cf. Fig. 21).

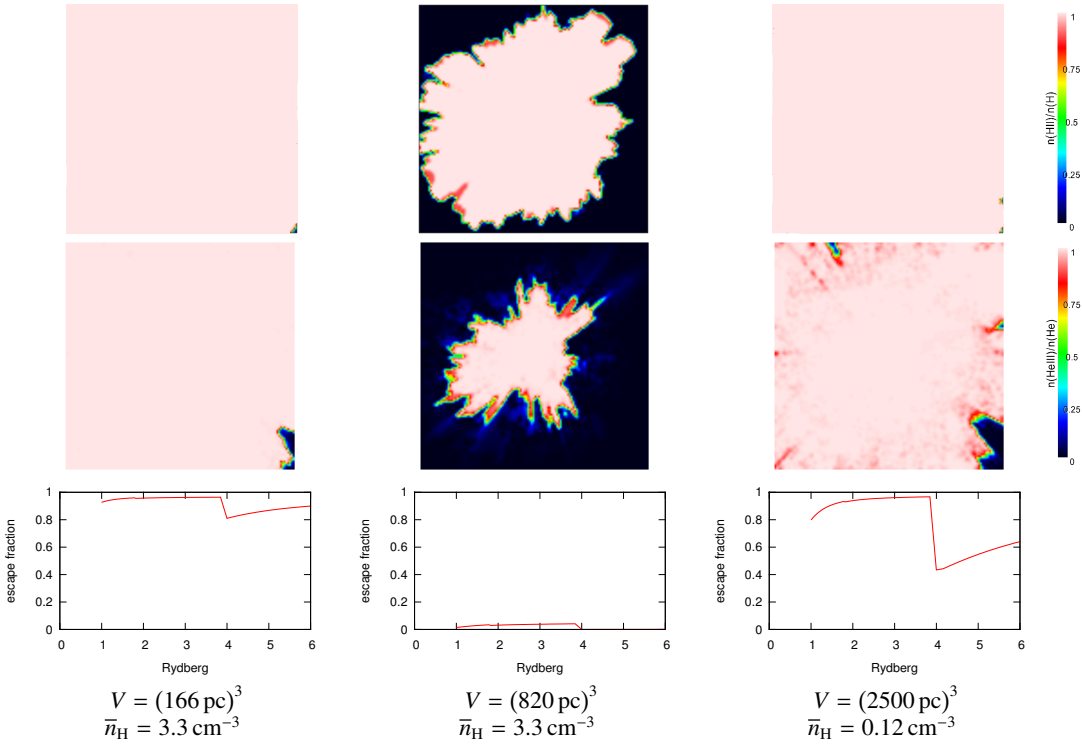

**Fig. 22.** Relative abundances of H II (top row) and He III (second row) for fractal density structures with the specified mean hydrogen number densities $\bar{n}_H$ and volumes $V$. The third row shows the calculated escape fractions as a function of photon energy for each of the simulations. While the escape fraction represents the "matter bounded" case in the left hand side model ($f_{esc} = 1$), the center model represents the "radiation bounded" case ($f_{esc} = 0$). In all cases the escape fraction for He II-ionizing photons is smaller than the escape fraction for H I-ionizing photons.

The most important aspect of the inhomogeneous structure of the ISM is its "porosity", and the characteristic parameter relevant for cosmic reionization simulations is the fraction of emitted ionizing photons that can escape the local environment and reach the intercluster medium and thus affect the reionization process. This so-called escape fraction is defined as the ratio of the number of photons not absorbed in a specified local reference volume to the total number of emitted photons. In order to calculate escape fractions for realistic scenarios of density distributions surrounding ionizing sources we have constructed a number of fractal density structures as proposed by Wood et al. (2005) (see also Elmegreen & Falgarone 1996). These structures are meant to represent the even smaller scales of star formation regions and evolved starburst clusters, as, for example, NGC 3603 and its surrounding thin ionized gas.

Fig. 22 shows the H II and He III ionization structures and the escape fractions as functions of the photon energy for three different representative models. In each case the central radiation source emits $8.5 \cdot 10^{52}$ H I-ionizing and $4.0 \cdot 10^{51}$ He II-ionizing photons per second (corresponding to a blackbody with a temperature of 95 000K). With a mean hydrogen number density of $3.3 \, \text{cm}^{-3}$ the first model corresponds to an H II region associated with star-formation clouds in the local universe (a volume of $(166 \, \text{pc})^3$, somewhat larger than the extension of the Carina nebula with a diameter of approximately 100 pc, cf. Smith et al. (2000), but smaller than NGC 604, cf. Melnick 1980). In this simulation, both hydrogen and helium are almost completely ionized and the majority of both H I-ionizing and He II-ionizing photons are able to escape from the simulated volume. In the next simulation the mean gas density has been retained, but the volume has been extended to $(820 \, \text{pc})^3$, which is approximately twice the size of NGC 604. In this case, less than 5 % of the H I-ionizing and almost none of the He II-ionizing photons are able to escape the volume. The last model has a mean hydrogen number density of $0.12 \, \text{cm}^{-3}$ and a volume of $(2.5 \, \text{kpc})$. This model approximates a typical spiral arm surrounded by the thin diffuse ionized gas (DIG) that in present-day disk galaxies contains a large fraction of the interstellar ionized gas (cf. Greenawalt et al. 1998, Hoffmann et al. 2012 and references therein). In this simulation the gas is almost transparent for soft H I-ionizing radiation, but it absorbs approximately 50 % of the He II-ionizing radiation (the escape fraction depends on the photon energy and reaches its minimum for photons whose energy is just above the ionization edge of He II). Within the energy intervals defined by the ionization edges, the frequency-dependent ionization cross section (cf. Sect. 2.2.2) leads to a radiation hardening, i.e. the escape fraction rises with larger photon energies. We note that the numerical values of the escape fractions can vary strongly even for small changes of the density structure and the properties of the ionizing sources, such that our results are just estimates for the escape fractions.

### 4.2.2. Simulations for the second stage of reionization using SEDs of massive and very massive stars

In this section we will focus on a high spectral resolution description of the 3d radiative transfer applied to the evolution of the ionization structures in the second stage of the reionization process, which was completed for hydrogen by a redshift of $z \sim 6$ and for helium by a redshift of $z \approx 3$ (cf. Sect. 1). As sources of ionization we consider metal-enriched population II stars, which had already come into being during this epoch. Although these objects are in general significantly cooler and less massive than the population III stars, they are numerous and powerful due to the strongly increased star for-mation rate during this epoch (Lineweaver 2001, Barger et al. 2000). Furthermore, there are indications that these stellar populations may have had top-heavy-IMFs which additionally generated VMSs (cf. Sect. 1).

With regard to reionization we simulate a volume of $(5.3\,\text{Mpc})^3$ in comoving units and assume (as in our simulations for the first stage of reionization) a homogeneous gas density corresponding to the mean baryonic density of the universe at this epoch. Although these approximations are quite restrictive (cf. Sect 4.2.1) we are convinced that the principal circumstances that would allow the much less massive population II and population I stars to fully reionize at least the hydrogen content of the IGM can be investigated in their essence with this approach.

Scenario based on stars representing the top of a Salpeter IMF. In this first step we combine the calculated SEDs of massive stars (cf. Fig. 10) with our time-dependent 3d radiative transfer simulations. The ionizing sources in these simulations represent clusters of stars, and their total hydrogen-ionizing photon emission rates are modelled on basis of a cluster with a Salpeter IMF. We have, however, approximated the spectral energy distribution of the cluster by the SEDs of its most luminous constituents, which we take to be $125\,M_\odot$ stars.

Depending on their evolution, galaxies and even different star forming regions within a single galaxy may have significantly different metallicities of their gas and, as a consequence, also of their young stellar populations[28]. We have therefore not used the same SED for all sources but instead use that of a $125\,M_\odot$ star with a metallicity of $Z = 0.1\,Z_\odot$ for 5 of the 10 clusters, and that of a $Z = 1.0\,Z_\odot$ model for the others. Based on our simulations in Sect. 4.2.1 we have chosen an escape fraction of $f_\text{esc} = 0.1$ as a plausible value for stars embedded in H II regions.

To model the luminosity of the clusters, we have used the population synthesis code Starburst99 (Leitherer et al. 1999, Leitherer et al. 2010) in continuous star formation mode on the basis of the mean cosmological star formation rate at the respective redshift (cf. Lineweaver 2001, Barger et al. 2000). The total luminosity of our sources (measured by their hydrogen-ionizing photon emission rates) was chosen to match the emission rates predicted by the Starburst99 model[29] for the corresponding volume. In the redshift range $10 > z > 6$ we assume a star formation rate of $0.02\,M_\odot\,\text{Mpc}^{-3}\,\text{yr}^{-1}$, in agreement with the mean value given for that interval by Lineweaver (2001), resulting in an emission rate of $4.7 \cdot 10^{53}$ H-ionizing and $1.0 \cdot 10^{50}$ He II-ionizing photons per second. For $6 > z > 3$ we increased the star formation rate to $0.05\,M_\odot\,\text{Mpc}^{-3}\,\text{yr}^{-1}$, leading to an enhanced emission of ionizing photons (H-ionizing flux of $1.2 \cdot 10^{54}$ photons per second and He II-ionizing flux of $2.5 \cdot 10^{50}$ photons per second). As shown in Fig. 23 the radiation of these cluster populations is able to reionize the hydrogen content of the simulation volume within the redshift interval from $z \sim 10$ to $z \sim 5.8$.

For stars with solar metallicity the He II-ionizing flux is reduced by more than four decades compared to stars with 0.1 solar metallicity (cf. Sect. 4.1) while there is no large variation of the H-ionizing flux. Only 1.4% of the He II atoms are ionized by the stellar populations in this simulation up to the redshift of $z \sim 2.8$ (Fig. 24) for which observations indicate that the reionization of He was completed (Reimers et al. 1997) – we note that replacing the spectra of the $Z = Z_\odot$ clusters by SEDs of low-metallicity (e.g. $Z = 0.1\,Z_\odot$) stars would not help much, since this would increase the fraction of He III only by a factor of less than 2, when compared to the "mixed" population presented in this section. Thus, it appears plausible that stellar populations with Salpeter-like IMFs are the main contributors to the reionization process of hydrogen, but due to the low emission rates of He II-ionizing photons they cannot contribute significantly to the ionization of He II.

Scenario based on stars representing runaway collision mergers. As the reionization of He II was considerably delayed compared to the reionization of H I, it appears likely that the sources of ionization responsible for the reionization of He II are different from those which caused the reionization of H and neutral He (cf. Sect. 1). Runaway collision mergers clearly exceed the assumed upper limit of the mass of present-day massive stars, but they are 1 to 2 orders of magnitude more efficient in producing He III-ionizing photons than normal hot stars based on a standard IMF, and since these stars can appear especially in chemically evolved clusters of high core mass density, thus at the temporal beginning of the star bursts, they may play an important role in these late stages of the reionization history.

To compare the efficacy of collisional mergers to that of "normal" stars with regard to the reionization of He, our second simulation uses the same star formation rate as before ($0.02\,M_\odot\,\text{Mpc}^{-3}\,\text{yr}^{-1}$ for $10 > z > 6$ and $0.05\,M_\odot\,\text{Mpc}^{-3}\,\text{yr}^{-1}$ for $z < 6$), but instead of a Salpeter IMF we assume that a fraction of 1% of the newly formed stellar mass is involved in mergers to become very massive stars[30]. (This assumption is based on the observationally determined ratio between the central black hole mass and the stellar mass in galaxies with active galactic nuclei at redshifts of $z \sim 4$ (Targett et al. 2012) as well as for the redshift range $1 < z < 2$ (Bennert et al. 2011) which is in the order of 1 %.[31]) For this simulation we approximate the cluster SED by that of a $3000\,M_\odot$ star with a metallicity of $Z = 0.05\,Z_\odot$ and, given that these objects are not created directly from gas clouds, but are the product of merging processes of already-formed stars, we assume an escape fraction of $f_\text{esc} = 0.5$ on the basis that these objects are able to remove the gas from their environment more quickly than ordinary massive stars due to their extremely strong winds (Pauldrach et al. 2012). The lifetime of such a very massive star is $2 \cdot 10^6$ yr (Pauldrach et al. 2012), resulting in a mean number of 20 VMSs within the simulated volume at any given time for the redshift range $10 > z > 6$ (with a total emission of $1.5 \cdot 10^{53}$ H-ionizing photons per second and $1.2 \cdot 10^{51}$

---

[28] E.g., the Small Magellanic Cloud has a metallicity of 1/6 solar (Dufour 1984), while the metallicity of M 83 is at least twice solar (Dufour et al. 1980, Rubin et al. 2007).

[29] The Starburst99 model uses a modified Salpeter IMF ("Kroupa IMF", Kroupa 2001) with an upper limit of $125\,M_\odot$ and a lower limit of $0.1\,M_\odot$ and two different slopes: $\Gamma = -0.3$ for $0.1\,M_\odot \leq M < 0.5\,M_\odot$ and $\Gamma = -1.3$ for $0.5 \leq M \leq 125$. The emergent fluxes refer to a model for a population with solar metallicity.

[30] Collissional mergers in dense starforming regions have been simulated on basis of $N$-body simulations by Portegies Zwart et al. (1999), Portegies Zwart & McMillan (2002), Gürkan et al. (2004), and Freitag et al. (2006). The properties of merger products have further been investigated by Suzuki et al. (2007) who find that neither the mass loss by the merging process nor the stellar winds are sufficient to inhibit the growth process of the runaway merger products in dense clusters. Very massive stars are therefore possible progenitors of the IMBHs (cf. Belkus et al. 2007, Pauldrach et al. 2012) that in turn might form the SMBHs found in the centers of galaxies (Ebisuzaki et al. 2001).

[31] For the present-day universe Magorrian et al. (1998) found, analyzing a sample of 32 nearby galaxies, that the mass of dark massive objects in the centers of galaxies (which are likely to be black holes) is approximately 0.6 % of the stellar mass within the galactic bulges in the corresponding host galaxies. Also, a similar mass ratio (2 %) between a central black hole and the stellar content has been found for the globular cluster $\omega$ Centauri (Noyola et al. 2010).

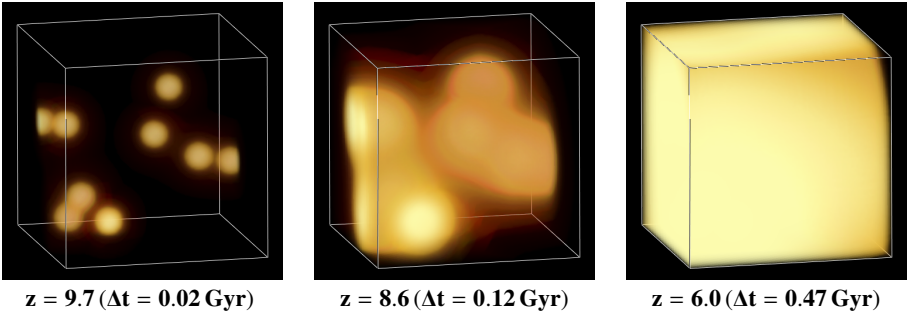

**Fig. 23.** Simulation representing the second stage of the cosmic reionization, showing the evolution of the hydrogen ionization fronts in the IGM. The volume is $(5.3\,\mathrm{Mpc})^3$ in comoving units and contains ten randomly distributed clusters of massive population II/I stars following a Kroupa IMF. The radiation of the clusters is able to reionize the hydrogen content of the simulation volume within the redshift interval from $z \sim 10$ to $z \sim 5.8$. ($\Delta t$ is the time interval since the beginning of the simulation at $z = 10$.)

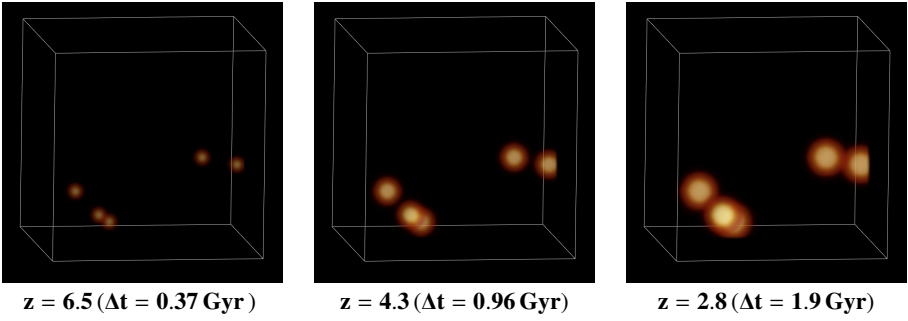

**Fig. 24.** As Fig. 23, but showing He III. At a redshift of $z \sim 2.8$ (for which observations indicate that the reionization of He was completed) only 1.4% of helium is ionized to He III in this simulation. Thus, due to their low emission rates of He II-ionizing photons, stellar populations with Salpeter-like IMFs obviously cannot contribute significantly to the cosmic reionization of helium.

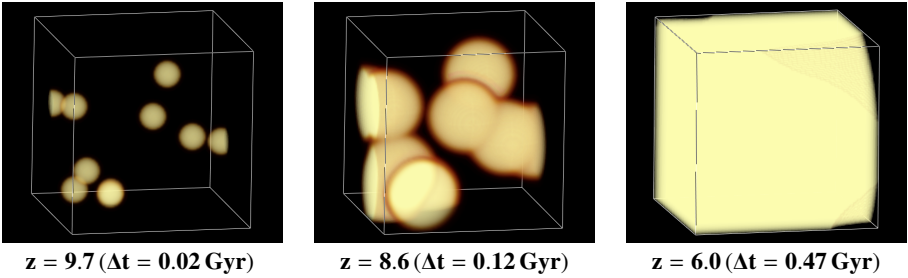

**Fig. 25.** As Fig. 23, but with clusters dominated by very massive stars (3000 $M_\odot$). These clusters are also able to reionize the hydrogen content of the simulation volume within the redshift interval from $z \sim 10$ to $z \sim 5.7$.

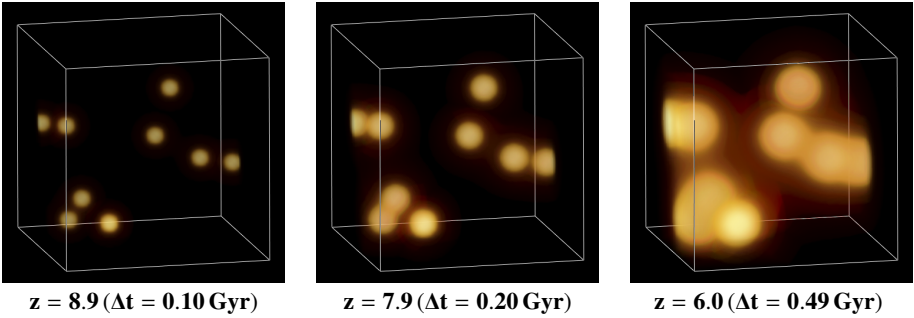

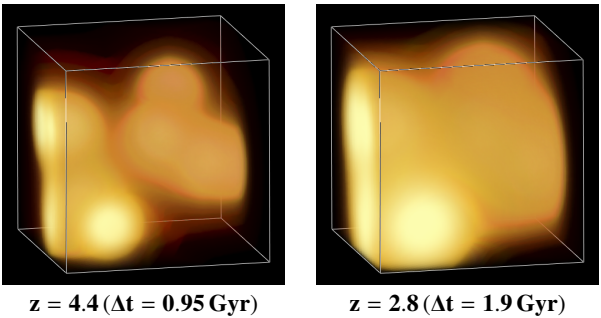

**Fig. 26.** As Fig. 25, but showing He III. In contrast to our simulation using normal stellar populations (Fig. 24) the clusters dominated by very massive stars can ionize the helium content of the simulation volume within approximately 2 Gyr, completing the reionization at a redshift of about $z \sim 2.5$. Thus, this scenario based on VMSs as the possible progenitors of IMBHs and SMBHs offers an alternative picture for the as yet unexplained cosmic reionization of helium. In these simulations recombination is not entirely negligible, in particular for helium, where the calculations including recombination yield a number of ionized particles about a factor of two smaller than it would yield for a calculation without recombination.

He II-ionizing per second) and 50 VMSs for the redshift range $6 > z > 2.5$ (with a total emission of $3.8 \cdot 10^{53}$ H-ionizing and $2.9 \cdot 10^{51}$ He II-ionizing photons per second).

With regard to hydrogen this scenario is similar to the one using the Salpeter IMF, resulting in a reionization completed at a redshift of $z \sim 5.7$ (cf. Fig. 25). For helium however, the VMSs are much more successful, producing 60% of the photons required to fully ionize He II up to a redshift of $z \sim 2.8$ (Fig. 26). As the figure clearly shows, this scenario based on the possi-ble progenitors of IMBHs and SMBHs offers an alternative or at least complementary picture for the pending explanation of the reionization of He.

## 5. Summary, conclusions, and outlook

The impact of massive stars on their environment is of major importance not only for the evolution of most galaxies but also for the ionization balance of the IGM. Although rare by num-ber, massive stars dominate the life cycle of gas and dust in star forming regions and are responsible for the chemical enrichment of the ISM and IGM. This is mainly due to the short lifetimes of massive stars, which favors the recycling of heavy elements on short timescales. Furthermore, since massive stars mostly group in young clusters, the combined amount of momentum and energy released by these clusters into the ISM controls the dynamical evolution of the gas, leading to the formation of superbubbles and triggered continuous star formation. The evolutionary behavior of these starbursts is obviously controlled by the metallicity, as is the steepness of the IMF, which can be more top-heavy than previously thought even in present-day galaxies. A common property among starbursts showing such flat IMFs is a high stellar density in the core of the clusters, and very recently strong evidence has been found that in these cases nature favors the formation of massive stars with masses $> 100 M_\odot$ (Harayama et al. 2008, Crowther et al. 2010). Moreover, during the evolution of dense stellar clusters runaway collision mergers can occur, leading to the formation of VMSs (objects of several 1000 $M_\odot$) *even in chemically evolved clusters* (cf. Belkus et al. 2007, Pauldrach et al. 2012), and therefore runaway collision mergers can play an important role in the late stages of the reionization history of the universe, whose detailed description is still one of the key unknowns in present cosmological research.

It is known meanwhile that the first generations of stars have not been the most important contributors to this process, since there is evidence that the universe was reionized in two steps (Cen 2003). The first one occurred – primarily driven by population III stars – up to a redshift of $z \sim 15$, and after a second cosmological almost-recombination phase (around $\Delta z \sim 5$, see Cen 2003) the reionization was driven by other powerful ionizing sources and completed at $z \sim 6$. Along with an increasing metallicity and star formation rate in this epoch (Lineweaver 2001, Barger et al. 2000), a top-heavy-IMF and VMSs – which are efficient emitters of ionizing photons with approximately 10 times more hydrogen-ionizing photons and 10 000 times more He II-ionizing photons per unit stellar mass than stars with a Salpeter IMF (cf. Pauldrach et al. 2012) – are determining factors for this process.

In this paper we have investigated whether such an extended view of the evolution of the universe from its dark ages to its final reionized state leads to a better understanding of the reionization process. For this purpose we have developed a 3-dimensional radiative transfer code (based on a ray-tracing method) in order to describe the evolution of the ionization fronts at different length scales. The main focus of this code is a comprehensive description of the time-dependent ionization structure with regard to a high spectral resolution of the ionizing spectra in order to accurately quantify the evolution of the ionization structures of different elements. Although only H and He are treated in the present paper, our algorithm allows the implementation of multiple levels for each ionization stage of the metals, and we are currently working on the inclusion of all important elements as described by Hoffmann et al. (2012). (As metal cooling has a significant influence on the temperature structure of the gas and therefore on the temperature dependent recombination rates and thus on the ionization structure of the most abundant elements, the temperature structure has to be treated consistently.) The geometrical aspects and the evolution of the ionization fronts have been extensively tested by comparisons to analytical models, the results of radially symmetric calculations, and tests presented by the "Cosmological Radiative Transfer Comparison Project" (Iliev et al. 2006). We found good agreement in all of these cases, showing that the important physical mechanisms which control the temporal expansions of the ionization fronts in homogeneous and inhomogeneous gas structures surrounding numerous sources of ionization have been implemented accurately.

For a realistic modelling of ionization fronts not only the structure of the ISM and IGM has to be correctly considered, but also the spectral energy distributions (SEDs) of the ionizing sources used as input for the radiative transfer. Thus, based on a sophisticated model atmosphere code which treats the physics of expanding stellar atmospheres consistently, an accurate simulation of realistic SEDs from population III to population I stars at various metallicities, temperatures, and masses has been an important aspect of our work. As different characteristics of the SEDs of the sources generally have a strong impact on the power of ionization, the usual application of oversimplified spectra (such as blackbodies) can lead to incorrect or misleading results with respect to the thickness and the extent of the ionization fronts in radiative transfer codes. We have *quantified* such possible systematic errors in the context of hot massive stars via a comparison of blackbody and realistic SEDs we have calculated at different metallicities. The results we present for the radial ionization structure of hydrogen and helium surrounding a $125 M_\odot$ star with different metallicities show that the variation of the metallicity of the star does not have a huge influence on the number of *hydrogen*-photons. However, this result does not reflect the influence of the metals in the gas which have been neglected up to now. A consistent computation including metals will actually lead to smaller temperatures for larger metallicities and thus result in smaller Strömgren spheres; then, an even larger influence on the ionization structure of the metals in the gas and thus on the gas temperatures and Strömgren spheres will be caused by non-LTE effects acting in the expanding atmospheres of the stars and determining the shapes of the SEDs, since in this case not only the number of ionizing photons is decisive but also their spectral distribution. Regarding the radial density structure of He III, the differences of the Strömgren radii which result from different metallicities of the ionizing sources are already considerably more pronounced due to this reason. This result clearly shows that the number of He II-ionizing photons is not just strongly influenced by the effective temperature and hence the mass of the objects, but also by the evolution of the metallicity. Employing blackbody spectra in general therefore leads to – at least in the case of helium and metals whose ionization energies exceed those of H – oversimplified ionization structures which do not represent the real structures. As exactly those elements which are affected by this approximation produce the observable emission lines, such a weakness in the simulation of the ionization structures would make a comparison to forthcoming observations questionable.

As a central application of our models with high spectral resolution we considered a number of different physical scenarios leading to ionizing sources with corresponding characteristics of the SEDs, which we compared to analogous calculations carried out with blackbodies. In particular we considered massive stars with 125 solar masses, representing the top of a Salpeter IMF, and very massive stars with 3 000 solar masses which are expected to form as runaway collision mergers in cluster-core collapses. Due to the large luminosities and effective temperatures of the latter objects the emitted number of H I-ionizing photons is, compared to stars at the top of a Salpeter IMF, larger by a factor of 40. This leads to hydrogen Strömgren radii around VMSs which are a factor of 3.5 larger than those around normal massive stars at the top of a Salpeter IMF. The effect is even more pronounced for He III where the VMS-Strömgren radii exceed those of the $125\,M_\odot$ stars by a factor of up to 100. The reason for this behavior is of course the ratio of the ionizing fluxes blueward of the He II-threshold, which favors very massive stars by a factor of millions. Very massive stars could thus strongly influence the ionization structure of their environment via their distinct tendency of the emission of H and He ionizing photons.

In the context of the first stage of the reionization history we have shown results of representative 3d radiative transfer simulations which are based on clusters of massive population III stars surrounded by a primordial environment reflecting a homogeneous intergalactic cosmological gas at $z \sim 15$. The calculations showed that our population III stars were able to reionize H as well as He on a timescale of just $t \sim 50$ Myr (corresponding to $\Delta z \sim 1.5$ at $z = 15$) within our chosen volume of $(5.3\,\text{Mpc})^3$. As the assumption of a homogeneous universe is just a rough approximation of the real situation, we have further calculated models based on inhomogeneous density structures, simulating volumes that comprise multiple ionizing sources on smaller scales. The "clumpiness" of the gas (including the star-forming density peaks of the ISM) results in considerably more complex ionization structures than those of models with homogeneous gas distributions, and as a consequence the fraction of ionizing photons which can escape their local environment and reach the intercluster medium can differ significantly. To estimate these escape fractions we have modelled inhomogeneous environments with fractal density structures of different sizes and mean densities ranging from the density of star-forming H II regions to the one of the diluted interstellar gas. In the simulations the escape fractions for the ionizing photons, and thus their ability to leak into the IGM and contribute to the ionization process of the universe, depend crucially on the size of the density peaks the ionizing sources are embedded in, ranging from less than 5 % for a very large and therefore radiation-bounded H II region, to almost 100 % for a smaller, matter-bounded H II region where the hydrogen content is almost completely ionized and the gas is therefore transparent to radiation. We note that even small variations in the number of absorbers can lead to strong changes of the escape fractions; thus, a precise determination of this value would require knowledge of the statistics of the density distribution in the small-scale environment of the sources (i.e., on the order of parsecs to a few kiloparsecs). Furthermore, for a given density distribution the escape-fractions are a function of the photon energy, and reach their minimum just blueward of the ionization edge of He II. The spectral energy distribution of the ionizing radiation escaping into the IGM may thus be different from the intrinsic SEDs of the stellar sources.

Although the initial conditions with respect to the structure and the content of the clusters are not yet precisely known, the result that population III stars have been able to reionize H and He within the generally accepted time frame is not spectacular, since it just reflects the current state of knowledge. The more interesting question regards the second stage of reionization – which completed at a redshift of $z \sim 6$ (for H) resp. at a redshift of $z \sim 2.8$ (for He) –, since it is by no means clear whether the metal-enriched and thus significantly cooler and less massive population II and population I stars, given a reasonable cluster structure and composition, have been able to fully reionize the hydrogen and helium content of the IGM. To investigate this scenario we have combined our predicted SEDs of massive and very massive stars with our time-dependent 3d radiative transfer simulations of the IGM. (Note that while the assumption of blackbody sources is not too bad an approximation for stars of primordial composition, it is not an adequate description of the real SEDs of later stellar populations. For the study of the second stage of reionization the results of realistic model atmospheres – as presented here – must therefore serve as input spectra.)

Based on plausible numbers for a standard population we have approximated in a first step the SED of a cluster with a Salpeter IMF by the SED of its most luminous members, namely $125\,M_\odot$ stars. Our calculations show that these objects have certainly been able to reionize H again (for our chosen volume the process required a time scale of $\Delta t \sim 500$ Myr, corresponding to the redshift interval between $z \sim 10$ and $z \sim 6$), but not He II (not even up to a redshift of $z \sim 2$ – note that the observations show that the reionization of He II was considerably delayed compared to the reionization of H). This latter finding is not unexpected, given that other investigations (Wyithe & Loeb 2003, Gleser et al. 2005, and McQuinn et al. 2009) had concluded from similar considerations that the sources of ionization responsible for the reionization of He II are different from those which caused the reionization of H.

To investigate whether the appearance of very massive, already metal-enriched population II and population I stars could have been responsible for the reionization of He II we replaced the $125\,M_\odot$ stars by runaway collision mergers of $M = 3\,000\,M_\odot$, which clearly exceed the assumed upper mass limit of present-day massive stars and which are up to 2 orders of magnitude more efficient in producing ionizing photons than normal hot stars based on a standard IMF are. The result is striking: In contrast to the case where the stellar populations follow a Salpeter or Kroupa IMF our simulations now show that the universe could have been reionized within approximately 2 Gyr (corresponding to a redshift of $z \sim 2.5$) in an environment where a fraction of 1 % of the mass formed into stars consists of mergers resulting in very massive stars. We note that this percentage is consistent with the mass ratio obtained for IMBHs and SMBHs relative to the stellar mass found in active galaxies and globular clusters like $\omega$ Centauri, and that such VMSs are possible progenitors of IMBHs and SMBHs. This result therefore offers an alternative or complementary scenario for the explanation of the reionization of He II to He III, and implies that a more realistic simulation of the entire reionization process of the universe must take into account the complete set of stellar populations and not just the primordial generation of stars.

It will therefore be necessary to understand the chemical evolution and the structure formation at the corresponding epochs more precisely and to establish comprehensive grids of realistic SEDs representing the time dependent content of the clusters and the cluster structures. Our future work will further focus on a more sophisticated description of the 3d radiative transfer with respect to high spectral resolution in order to quantify the evolution of all relevant ionization structures accurately. On basis of synthetic emission spectra calculated for the 3d structure of the ionized gas we will try to determine possible observational features (such as line strength ratios from different ions) that may be used to discriminate between the details of different reionization scenarios. Knowledge of such features will be important for comparisons with more detailed observations at high redshifts, which are expected to become available during the next years.

*Acknowledgements.* We thank an anonymous referee for helpful comments which improved the paper and we further wish to thank I. Iliev for granting us access to the data of the *Cosmological Radiative Transfer Codes Comparison* and N. Gnedin for providing us with the IFrIT program used for some visualizations of our results. The calculations were performed on high-performance computing facilities at the LRZ (Leibniz Rechenzentrum) and the University of Munich which is to be acknowledged. This work was supported by the *Deutsche Forschungsgemeinschaft (DFG)* under grant PA 477/9-1 and PA 477/18-1.

## References

Abel, T., Bryan, G. L., & Norman, M. L. 2000, ApJ, 540, 39
Abel, T., Norman, M. L., & Madau, P. 1999, ApJ, 523, 66
Abel, T. & Wandelt, B. D. 2002, MNRAS, 330, L53
Alvarez, M. A., Bromm, V., & Shapiro, P. R. 2006, ApJ, 639, 621
Baker, J. G. & Menzel, D. H. 1938, ApJ, 88, 52
Barger, A. J., Cowie, L. L., & Richards, E. A. 2000, AJ, 119, 2092
Belkus, H., Van Bever, J., & Vanbeveren, D. 2007, ApJ, 659, 1576
Bennert, V. N., Auger, M. W., Treu, T., Woo, J.-H., & Malkan, M. A. 2011, ApJ, 742, 107
Bromm, V., Coppi, P. S., & Larson, R. B. 2002, ApJ, 564, 23
Caffau, E., Bonifacio, P., François, P., et al. 2012, A&A, 542, A51
Carroll, B. W. & Ostlie, D. A. 2006, An introduction to modern astrophysics and cosmology, ed. Carroll, B. W. & Ostlie, D. A.
Cen, R. 2003, ApJ, 591, 12
Ciardi, B., Ferrara, A., Marri, S., & Raimondo, G. 2001, MNRAS, 324, 381
Clark, P. C., Glover, S. C. O., Smith, R. J., et al. 2011, Science, 331, 1040
Crowther, P. A., Schnurr, O., Hirschi, R., et al. 2010, MNRAS, 408, 731
Dufour, R. J. 1984, in IAU Symposium, Vol. 108, Structure and Evolution of the Magellanic Clouds, ed. S. van den Bergh & K. S. D. de Boer, 353–360
Dufour, R. J., Talbot, Jr., R. J., Jensen, E. B., & Shields, G. A. 1980, ApJ, 236, 119
Ebisuzaki, T., Makino, J., Tsuru, T. G., et al. 2001, ApJ, 562, L19
El Eid, M. F., Fricke, K. J., & Ober, W. W. 1983, A&A, 119, 54
Elmegreen, B. G. & Falgarone, E. 1996, ApJ, 471, 816
Ercolano, B., Barlow, M. J., Storey, P. J., & Liu, X.-W. 2003, MNRAS, 340, 1136
Fan, X., Carilli, C. L., & Keating, B. 2006a, ARA&A, 44, 415
Fan, X., Narayanan, V. K., Lupton, R. H., et al. 2001, AJ, 122, 2833
Fan, X., Strauss, M. A., Richards, G. T., et al. 2006b, AJ, 131, 1203
Freitag, M., Gürkan, M. A., & Rasio, F. A. 2006, MNRAS, 368, 141
Furlanetto, S. R. & Loeb, A. 2005, ApJ, 634, 1
Gal-Yam, A., Mazzali, P., Ofek, E. O., et al. 2009, Nature, 462, 624
Gleser, L., Nusser, A., Benson, A. J., Ohno, H., & Sugiyama, N. 2005, MNRAS, 361, 1399
Gnedin, N. Y. 2000, ApJ, 535, 530
Gnedin, N. Y. & Abel, T. 2001, New Astronomy, 6, 437
Gnedin, N. Y. & Bertschinger, E. 1996, ApJ, 470, 115
Gnedin, N. Y. & Ostriker, J. P. 1997, ApJ, 486, 581
Górski, K. M., Hivon, E., Banday, A. J., et al. 2005, ApJ, 622, 759
Graziani, L., Maselli, A., & Ciardi, B. 2012, ArXiv e-prints
Greenawalt, B., Walterbos, R. A. M., Thilker, D., & Hoopes, C. G. 1998, ApJ, 506, 135
Gunn, J. E. & Peterson, B. A. 1965, ApJ, 142, 1633
Gürkan, M. A., Freitag, M., & Rasio, F. A. 2004, ApJ, 604, 632
Harayama, Y., Eisenhauer, F., & Martins, F. 2008, ApJ, 675, 1319
Hernquist, L. & Springel, V. 2003, MNRAS, 341, 1253
Hoffmann, T. L., Lieb, S., Pauldrach, A. W. A., et al. 2012, A&A, 544, A57
Iliev, I. T., Ciardi, B., Alvarez, M. A., et al. 2006, MNRAS, 371, 1057
Iliev, I. T., Whalen, D., Mellema, G., et al. 2009, MNRAS, 400, 1283
Jelic, V. 2010, PhD thesis, Kapteyn Astronomical Institute, University of Groningen,vjelic@astro.rug.nl
Kashikawa, N. 2007, in Astronomical Society of the Pacific Conference Series, Vol. 380, Deepest Astronomical Surveys, ed. J. Afonso, H. C. Ferguson, B. Mobasher, & R. Norris, 11–+
Klessen, R. S., Glover, S. C. O., & Clark, P. C. 2012, MNRAS, 421, 3217
Kriss, G. A., Shull, J. M., Oegerle, W., et al. 2001, Science, 293, 1112
Kroupa, P. 2001, MNRAS, 322, 231
Leitherer, C., Ortiz Otálvaro, P. A., Bresolin, F., et al. 2010, ApJS, 189, 309
Leitherer, C., Schaerer, D., Goldader, J. D., et al. 1999, ApJS, 123, 3
Lineweaver, C. H. 2001, Icarus, 151, 307
Magorrian, J., Tremaine, S., Richstone, D., et al. 1998, AJ, 115, 2285
Mahabal, A., Stern, D., Bogosavljević, M., Djorgovski, S. G., & Thompson, D. 2005, ApJ, 634, L9
Maio, U., Ciardi, B., Dolag, K., Tornatore, L., & Khochfar, S. 2010, MNRAS, 407, 1003
Maio, U., Ciardi, B., Yoshida, N., Dolag, K., & Tornatore, L. 2009, A&A, 503, 25
McQuinn, M., Lidz, A., Zahn, O., et al. 2007, MNRAS, 377, 1043
McQuinn, M., Lidz, A., Zaldarriaga, M., et al. 2009, ApJ, 694, 842
Mellema, G., Iliev, I. T., Alvarez, M. A., & Shapiro, P. R. 2006, New Astronomy, 11, 374
Melnick, J. 1980, A&A, 86, 304
Mihalas, D. 1978, Stellar atmospheres /2nd edition/ (San Francisco, W. H. Freeman and Co., 1978. 650 p.)
Nakamoto, T., Umemura, M., & Susa, H. 2001, MNRAS, 321, 593
Norman, M. L. 2010, in American Institute of Physics Conference Series, Vol. 1294, American Institute of Physics Conference Series, ed. D. J. Whalen, V. Bromm, & N. Yoshida, 17–27
Noyola, E., Gebhardt, K., Kissler-Patig, M., et al. 2010, ApJ, 719, L60
Osterbrock, D. E. & Ferland, G. J. 2006, Astrophysics of gaseous nebulae and active galactic nuclei (Astrophysics of gaseous nebulae and active galactic nuclei, 2nd. ed. by D.E. Osterbrock and G.J. Ferland. Sausalito, CA: University Science Books, 2006)
Pauldrach, A. 1987, A&A, 183, 295
Pauldrach, A. W. A., Hoffmann, T. L., & Lennon, M. 2001, A&A, 375, 161
Pauldrach, A. W. A., Kudritzki, R. P., Puls, J., Butler, K., & Hunsinger, J. 1994, A&A, 283, 525
Pauldrach, A. W. A., Vanbeveren, D., & Hoffmann, T. L. 2011, ArXiv e-prints 1107.0654v2
Pauldrach, A. W. A., Vanbeveren, D., & Hoffmann, T. L. 2012, A&A, 538, A75
Petkova, M. & Springel, V. 2009, MNRAS, 396, 1383
Portegies Zwart, S. F., Dewi, J., & Maccarone, T. 2004, MNRAS, 355, 413
Portegies Zwart, S. F., Makino, J., McMillan, S. L. W., & Hut, P. 1999, A&A, 348, 117
Portegies Zwart, S. F. & McMillan, S. L. W. 2002, ApJ, 576, 899
Press, W. H., Teukolsky , S. A., Vetterling, W. T., & Flannery, B. P. 1992, Numerical recipes in FORTRAN. The art of scientific computing (Cambridge: University Press, —c1992, 2nd ed.)
Razoumov, A. O. & Cardall, C. Y. 2005, MNRAS, 362, 1413
Reimers, D., Kohler, S., Wisotzki, L., et al. 1997, A&A, 327, 890
Reynolds, D. R., Hayes, J. C., Paschos, P., & Norman, M. L. 2009, Journal of Computational Physics, 228, 6833
Ritzerveld, J., Icke, V., & Rijkhorst, E.-J. 2003, ArXiv Astrophysics e-prints
Rubin, R. H., Simpson, J. P., Colgan, S. W. J., et al. 2007, MNRAS, 377, 1407
Rubin, R. H., Simpson, J. P., Haas, M. R., & Erickson, E. F. 1991, ApJ, 374, 564
Ryu, D., Ostriker, J. P., Kang, H., & Cen, R. 1993, ApJ, 414, 1
Santos, M. G., Silva, M. B., Pritchard, J. R., Cen, R., & Cooray, A. 2011, A&A, 527, A93+
Schaerer, D. 2002, A&A, 382, 28
Schmidt-Voigt, M. & Köeppen, J. 1987, A&A, 174, 211
Schneider, R., Omukai, K., Limongi, M., et al. 2012, MNRAS, 423, L60
Sellmaier, F. H., Yamamoto, T., Pauldrach, A. W. A., & Rubin, R. H. 1996, A&A, 305, L37
Smith, N., Egan, M. P., Carey, S., et al. 2000, ApJ, 532, L145
Spitzer, L. 1998, Physical Processes in the Interstellar Medium (Physical Processes in the Interstellar Medium, by Lyman Spitzer, pp. 335. ISBN 0-471-29335-0. Wiley-VCH , May 1998.)
Suzuki, T. K., Nakasato, N., Baumgardt, H., et al. 2007, ApJ, 668, 435
Syphers, D., Anderson, S. F., Zheng, W., et al. 2011, ArXiv e-prints
Targett, T. A., Dunlop, J. S., & McLure, R. J. 2012, MNRAS, 420, 3621
Trac, H. & Cen, R. 2007, ApJ, 671, 1
Volonteri, M. & Gnedin, N. Y. 2009, ApJ, 703, 2113
White, R. L., Becker, R. H., Fan, X., & Strauss, M. A. 2003, AJ, 126, 1
Willott, C. J., Delfosse, X., Forveille, T., Delorme, P., & Gwyn, S. D. J. 2005, ApJ, 633, 630
Wise, J. H. & Abel, T. 2008, ApJ, 684, 1
Wise, J. H. & Abel, T. 2011, MNRAS, 414, 3458
Wood, K., Haffner, L. M., Reynolds, R. J., Mathis, J. S., & Madsen, G. 2005, ApJ, 633, 295
Wyithe, J. S. B. & Loeb, A. 2003, ApJ, 586, 693
Yoshida, N., Bromm, V., & Hernquist, L. 2004, ApJ, 605, 579
Zanstra, H. 1931, ZAp, 2, 1